\documentclass[aps,pre,twocolumn,longbibliography,unsortedaddress,floatfix]{revtex4-1}

\usepackage{amsmath,amssymb,amsfonts}
\usepackage{algorithmic}
\usepackage{graphicx} 
\usepackage{color} 
\usepackage{textcomp}

\usepackage{mathtools}
\usepackage{bm}
\usepackage{url}

\newcommand{\BA}{\bm{A}}
\newcommand{\BF}{\bm{F}}
\newcommand{\BG}{\bm{G}}
\newcommand{\BH}{\bm{H}}
\newcommand{\BJ}{\bm{J}}
\newcommand{\BX}{\bm{X}}
\newcommand{\BZ}{\bm{Z}}
\newcommand{\Bf}{\bm{f}}
\newcommand{\calB}{\mathcal{B}}
\newcommand{\calF}{\mathcal{F}}

\newcommand{\inner}[2]{#1 \cdot #2}
\newcommand{\average}[2]{\left\langle #1 \right\rangle_{#2}}
\newcommand{\functional}[3]{#1\left\{ #2,#3 \right\}}
\newcommand{\func}[2]{#1^{(#2)}}

\newcommand{\norm}[1]{\left\| #1 \right\|}
\newcommand{\pd}[2]{\frac{\partial #1}{\partial #2}}

\begin{document}
\title{Optimal coupling functions for fast and global synchronization of\\ weakly coupled limit-cycle oscillators}
\author{Norihisa Namura}
\thanks{Corresponding author. E-mail: namura.n.aa@m.titech.ac.jp}
\affiliation{Department of Systems and Control Engineering, Tokyo Institute of Technology, Tokyo 152-8552, Japan}

\author{Hiroya Nakao}
\affiliation{Department of Systems and Control Engineering, Tokyo Institute of Technology, Tokyo 152-8552, Japan}

\date{\today}


\begin{abstract}
We propose a method for optimizing mutual coupling functions to achieve fast and global synchronization between a pair of weakly coupled limit-cycle oscillators.
Our method is based on phase reduction that provides a concise low-dimensional representation of the synchronization dynamics of mutually coupled oscillators,
including the case where the coupling depends on past time series of the oscillators.
We first describe a method for a pair of identical oscillators and then generalize it to the case of slightly nonidentical oscillators.
The coupling function is designed in two optimization steps for the functional form and amplitude, 
where the amplitude is numerically optimized to minimize the average convergence time under a constraint on the total power.
We perform numerical simulations of the synchronization dynamics with the optimized coupling functions using the FitzHugh-Nagumo and R\"{o}ssler oscillators as examples.
We show that the coupling function optimized by the proposed method can achieve global synchronization more efficiently than the previous methods.
\end{abstract}

\maketitle


\section{Introduction}
\label{sec:Introduction}

Synchronized rhythms are universally found in the real world and often play important functional roles~\cite{Pikovsky2001synchronization,Strogatz2003sync}. 
Physiological examples include brain waves~\cite{stankovski2015coupling,stankovski2017neural}, 
circadian rhythms~\cite{circadian1,circadian2},
animal gaits~\cite{gait1,gait2}, 
heartbeats and breathing~\cite{Kralemann2013vivo},
synchronous flashing of fireflies~\cite{firefly1,firefly2}, 
and synchronized secretion of insulin from pancreatic beta cells~\cite{Winfree2001geometry,betacell}.
Synchronized rhythms are also applied in various fields of engineering, for example, 
to control the dynamics of robots such as salamander robots~\cite{Ijspeert2007swimming} and hexapod robots~\cite{Minati2018versatile}, 
human robot interactions~\cite{Mortl2014rhythm}, and networks of power generators~\cite{Motter2013spontaneous,Dorfler2012synchronization}.

Stable periodic dynamics of rhythmic systems are typically modeled as limit-cycle oscillators~\cite{Strogatz2015nonlinear}.
Synchronization of limit-cycle oscillators occurs when they are periodically perturbed or mutually coupled. 
When a periodic input is given to a limit-cycle oscillator, \textit{entrainment} or \textit{phase-locking} can occur~\cite{Kuramoto1984chemical,Pikovsky2001synchronization}. 
When two or more oscillators are coupled, they can exhibit mutual synchronization, in which their rhythms align with each other.

Phase reduction is a useful method for analyzing synchronization of weakly coupled limit-cycle oscillators~\cite{Kuramoto1984chemical,Hoppensteadt1997weakly,Winfree2001geometry,Brown2004phase,Ermentrout2010mathematical,Nakao2016phase,Ashwin2016mathematical,Kuramoto2019concept,Ermentrout2019recent}. 
By phase reduction, nonlinear multidimensional dynamics of a limit-cycle oscillator are reduced to a single-variable phase equation characterized by the natural frequency and phase response of the oscillator. 
It has widely been used to describe nontrivial synchronization dynamics of coupled oscillators.
The phase reduction method is also useful for designing desirable dynamics of limit-cycle oscillators.
For example, we recently proposed a method to design a limit-cycle oscillator with a desirable periodic orbit and phase-response characteristics~\cite{Namura2023designing}.

Recently, the phase reduction method has also been applied to the control of rhythmic systems to realize efficient synchronization dynamics.
Many studies have been conducted to find optimal inputs that realize desired entrainment dynamics under various conditions~\cite{Pyragas2018optimal,Monga2018synchronizing,monga2019phase1},
in particular, optimizing the input power~\cite{Moehlis2006optimal,Harada2010optimal,Dasanayake2011optimal,Zlotnik2012optimal,Wilson2015optimal,Monga2019optimal},
locking range~\cite{Tanaka2014optimal,Tanaka2015optimal,Yabe2020locking},
convergence time~\cite{Qiao2017entrainment,Kato2021optimization}, 
and linear stability of the phase-locking point~\cite{Zlotnik2013optimal,Takata2021fast}.
For a large population of oscillators, control of the phase distribution~\cite{Monga2018synchronizing,Kuritz2019ensemble,Monga2019phase2}
and phase-selective entrainment have also been performed~\cite{Zlotnik2016phase,Singhal2023optimal}.
For rhythmic systems perturbed by noise, maximization of the phase coherence has also been considered~\cite{Pikovsky2015maximizing}.
In addition, methods for controlling limit-cycle oscillators using delayed feedback have also been studied
~\cite{Novicenko2012phase,Novicenko2015delayed}.

Optimization of synchronization in networks of rhythmic systems, including chaotic dynamics, 
has been extensively studied in the literature~\cite{Nishikawa2006synchronization,Nishikawa2006maximum,Tanaka2008optimal,Nishikawa2010network,Yanagita2010design,Ravoori2011robustness,Yanagita2012design,Yanagita2014thermodynamic,Skardal2014optimal,Skardal2016optimal,Nakao2021sparse}.
Regarding mutual synchronization between a pair of limit-cycle oscillators, 
optimization of diffusive coupling to improve the linear stability of mutual synchronization has been considered in~\cite{Shirasaka2017optimizing},
and further extended to the case where the oscillators can interact not only through their present states but also through their past time series~\cite{Watanabe2019optimization}. 
The optimal delay time, coefficients for linear temporal filtering, and optimal response and driving functions have been derived 
under the assumption that the mutual coupling has a drive-response structure~\cite{Watanabe2019optimization}.

In this study, we propose a general method that does not rely on such a specific coupling structure for optimizing mutual coupling to achieve fast and global synchronization.
We optimize the average convergence time towards synchronized states for coupled pairs of identical and slightly nonidentical oscillators.
Our numerical simulations demonstrate that the mutual coupling function optimized by the proposed method achieves more efficient global synchronization than the previous methods in~\cite{Watanabe2019optimization}.

This paper is organized as follows.
We first introduce a mathematical model of weakly coupled oscillators and reduced phase equations in Sec.~\ref{sec:model}.
We then describe a method to optimize the mutual coupling for fast and global synchronization in Sec.~\ref{sec:methods}.
In Sec.~\ref{sec:results}, we demonstrate the performance of the proposed method by numerical simulations and compare the results with the previous methods.
Concluding remarks are given in Sec.~\ref{sec:conclusion}.
Appendices provide a review of the previous methods and details of the numerical optimization methods.


\section{Model of Mutual Synchronization}
\label{sec:model}

In this section, we first introduce a general mathematical model of coupled oscillators and reduced phase equations, generalizing the results in~\cite{Watanabe2019optimization}.


\subsection{A pair of weakly coupled oscillators}
\label{sec:model_1}

We consider a pair of weakly coupled limit-cycle oscillators with nearly identical properties, 
where they can interact mutually not only through their current states but also through their past time series.
We assume that the dynamics of the oscillators are described by the following functional differential equations:
\begin{align}
\begin{aligned}
\label{eq:model}
\dot{\BX}_{1}(t) = \BF_{1}(\BX_{1}(t)) + \varepsilon\functional{\tilde{\BH}_{1}}{\func{\BX_{1}}{t}(\cdot)}{\func{\BX_{2}}{t}(\cdot)}, \\
\dot{\BX}_{2}(t) = \BF_{2}(\BX_{2}(t)) + \varepsilon\functional{\tilde{\BH}_{2}}{\func{\BX_{2}}{t}(\cdot)}{\func{\BX_{1}}{t}(\cdot)},
\end{aligned}
\end{align}
where $\BX_{1,2}(t) \in \mathbb{R}^{N}$ is the state of each oscillator at time $t$, 
the overdot denotes the time derivative,
$\func{\BX_{1,2}}{t}(\cdot)$ represents the past time series of each oscillator (see below),
$\BF_{1,2}: \mathbb{R}^{N} \to \mathbb{R}^{N}$ is a sufficiently smooth vector field describing the dynamics of each oscillator,
and $0 < \varepsilon \ll 1$ is a sufficiently small parameter representing weakness of the mutual coupling.

We assume that the properties of the two oscillators are nearly identical,
that is, we can split the vector field $\BF_{1,2}$ into a common part $\BF$ and a small deviation $\varepsilon \Bf_{1,2}$ as
\begin{align}
\BF_{1,2}(\BX) = \BF(\BX) + \varepsilon \Bf_{1,2}(\BX),
\end{align}
where $\Bf_{1,2}$ is of $O(1)$.
We assume that the common part of the vector field $\BF$ has an exponentially stable limit-cycle solution,
$\tilde{\BX}_{0}(t) = \tilde{\BX}_{0}(t + T)$, whose period is $T$ and natural frequency is $\omega = 2\pi/T$.
We also assume that the vector field of each oscillator is only slightly perturbed by the deviations and the mutual coupling.

In Eq.~\eqref{eq:model}, each functional $\tilde{\BH}_{1,2}: \calF \times \calF \to \mathbb{R}^{N}$ represents a sufficiently smooth mutual coupling that depends on the past time series of $\BX_{1,2}$,
where $\calF$ is a function space consisting of the time series of length $T_{L} \geq 0$.
We use the standard notation of functional differential equations~\cite{Hale1977theory} to represent the time series of $\BX_{1,2}$ in the coupling functional $\tilde{\BH}_{1,2}$.
The symbol $\func{\BX_{1,2}}{t}(\cdot) \in \calF$ represents the time series of $\BX_{1,2}$ on the interval $[t - T_{L},t]$, 
whose value at each point of time is defined by
\begin{align}
\label{eq:func_t}
\func{\BX_{1,2}}{t}(s) = \BX_{1,2}(t + s)
\end{align}
for $s \in [-T_{L}, 0]$.
The symbol $(\cdot)$ in Eq.~\eqref{eq:model} means that $\tilde{\BH}_{1,2}$ is a functional, which depends not only on the values of $\BX_{1,2}$ at a certain moment of time but generally on the time series of $\BX_{1,2}$.
We omit this symbol hereafter unless necessary.


\subsection{Phase reduction}
\label{sec:model_2}

The dynamics of weakly coupled limit-cycle oscillators can be analyzed by coupled phase equations, 
approximately derived from the original mathematical model by the phase reduction method~\cite{Kuramoto1984chemical,Hoppensteadt1997weakly,Winfree2001geometry,Brown2004phase,Ermentrout2010mathematical,Nakao2016phase,Ashwin2016mathematical,Kuramoto2019concept,Ermentrout2019recent}.
For an exponentially stable limit-cycle solution $\tilde{\BX}_{0}(t)$ obeying the dynamics $\dot{\BX} = \BF(\BX)$, we can introduce an asymptotic phase function $\Theta(\BX):\calB \to [0,2\pi)$ into the basin $\calB \subseteq \mathbb{R}^{N}$ of the limit cycle satisfying $\inner{\nabla\Theta(\BX)}{\BF(\BX)} = \omega$,
where $\inner{\bm{a}}{\bm{b}} = \sum_{i=1}^{N} a_{i} b_{i}$ represents the scalar product of two vectors $\bm{a}$, $\bm{b} \in \mathbb{R}^{N}$.
Using the asymptotic phase, we can define the phase value $\theta$ of the state $\BX \in \calB$ of the oscillator by $\theta = \Theta(\BX)$. 
Then, the phase value increases constantly at the frequency $\omega$ as
\begin{align}
\dot{\theta}(t) = \frac{d}{dt}\Theta(\BX(t)) = \inner{\nabla\Theta(\BX(t))}{\BF(\BX(t))} = \omega,
\end{align}
where $0$ and $2\pi$ are considered identical.
The state on the limit cycle can be expressed as $\BX_{0}(\theta) = \tilde{\BX}_{0}(t = \theta/\omega)$ as a function of the phase $\theta$, 
where $\BX_{0}(\theta)$ is a $2\pi$-periodic function satisfying $\BX_{0}(\theta) = \BX_{0}(\theta + 2\pi)$.
As in Eq.~\eqref{eq:func_t}, we can express the time series $\func{\BX_{0}}{\theta}(\cdot)$ on the limit cycle as a function of $\theta$, whose value at each point of time is defined by
\begin{align}
\func{\BX_{0}}{\theta}(s) = \BX_{0}(\theta + \omega s)
\end{align}
for $s \in [-T_{L}, 0]$.
In what follows, the symbol $(\cdot)$ will be omitted and the time series on the limit cycle will be denoted as $\func{\BX_{0}}{\theta}$.

We now denote the phase of each limit-cycle oscillator as $\theta_{1,2} = \Theta(\BX_{1,2})$.
As the perturbation applied to the oscillator is sufficiently weak and of $O(\varepsilon)$, the state of each oscillator can be approximated as $\BX_{1,2}(t) = \BX_{0}(\theta_{1,2}(t)) + O(\varepsilon)$.
If this approximation is valid over the whole time interval $[t - T_{L},t]$, the time series can be approximated by 
\begin{align}
\func{\BX_{1,2}}{t}(s) = \func{\BX_{0}}{\theta_{1,2}(t)}(s) + O(\varepsilon) \quad (s \in [-T_{L}, 0])
\end{align}
and the smoothness of the functional $\tilde{\BH}$ implies
\begin{align}
\label{eq:reduce_H}
\functional{\tilde{\BH}_{1,2}}{\func{\BX_{1}}{t}}{\func{\BX_{2}}{t}} = \functional{\tilde{\BH}_{1,2}}{\func{\BX_{0}}{\theta_{1}(t)}}{\func{\BX_{0}}{\theta_{2}(t)}} + O(\varepsilon T_{L}).
\end{align}
We assume that $T_{L}$ is of $O(1)$ in what follows. 
Substituting Eq.~\eqref{eq:reduce_H} into \eqref{eq:model} and ignoring the errors of $O(\varepsilon^{2})$,
we can obtain the following reduced phase equations:
\begin{align}
\begin{aligned}
\dot{\theta}_{1}(t) ={}& \omega + \varepsilon \inner{\BZ(\theta_{1}(t))}{\Bf_{1}(\BX_{0}(\theta_{1}(t)))} \\
&+ \varepsilon \inner{\BZ(\theta_{1}(t))}{\functional{\tilde{\BH}_{1}}{\func{\BX_{0}}{\theta_{1}(t)}}{\func{\BX_{0}}{\theta_{2}(t)}}}, \\
\dot{\theta}_{2}(t) ={}& \omega + \varepsilon \inner{\BZ(\theta_{2}(t))}{\Bf_{2}(\BX_{0}(\theta_{2}(t)))} \\
&+ \varepsilon \inner{\BZ(\theta_{2}(t))}{\functional{\tilde{\BH}_{2}}{\func{\BX_{0}}{\theta_{2}(t)}}{\func{\BX_{0}}{\theta_{1}(t)}}},
\end{aligned}
\end{align}
which are correct up to $O(\varepsilon)$.
Here, $\BZ: [0,2\pi) \to \mathbb{R}^{N}$ is the phase sensitivity function~(PSF) of the limit cycle $\BX_{0}(\theta)$, defined by $\BZ(\theta) = \left. \nabla\Theta(\BX) \right|_{\BX = \BX_{0}(\theta)}$.
The PSF characterizes linear response of the phase $\theta$ to a weak input applied to the oscillator state $\BX_{0}(\theta)$ on the limit cycle. 
It can be obtained as the $2\pi$-periodic solution to the following adjoint equation~\cite{Brown2004phase,Ermentrout2010mathematical}:
\begin{align}
\omega \frac{d}{d\theta}\BZ(\theta) = -\BJ(\BX_{0}(\theta))^{\top}\BZ(\theta),
\end{align}
where $\BJ: \mathbb{R}^{N} \to \mathbb{R}^{N \times N}$ is a Jacobian matrix of $\BF$ at $\BX$, i.e., $\BJ(\BX) = \nabla\BF(\BX)$.
The PSF should satisfy a normalization condition $\inner{\BZ(\theta)}{d\BX_{0}(\theta)/d\theta} = 1$ so that the identity $\Theta({\bm X}_0(\theta)) = \theta$ holds~\cite{Brown2004phase,Ermentrout2010mathematical}.

Although the mutual coupling $\functional{\tilde{\BH}_{1,2}}{\func{\BX_{0}}{\theta_{1}}}{\func{\BX_{0}}{\theta_{2}}}$ is a functional of two series $\func{\BX_{0}}{\theta_{1}}$ and $\func{\BX_{0}}{\theta_{2}}$,
we can regard this functional as an ordinary function of $\theta_1$ and $\theta_2$,
because the functions $\func{\BX_{0}}{\theta_{1}}$ and $\func{\BX_{0}}{\theta_{2}}$ are uniquely determined only by the phase values $\theta_{1}$ and $\theta_{2}$, respectively, provided that $\BX_{0}$ is the closed orbit of the limit cycle.
Therefore, we can define the mutual coupling functions~(MCFs) $\BH_{1,2}(\theta_{1},\theta_{2})$ as
\begin{align}
\BH_{1,2}(\theta_{1},\theta_{2}) = \functional{\tilde{\BH}_{1,2}}{\func{\BX_{0}}{\theta_{1}}}{\func{\BX_{0}}{\theta_{2}}}.
\end{align}
We thus obtain the following pair of ordinary differential equations as the reduced phase equations:
\begin{align}
\begin{aligned}
\label{eq:phase_eq}
\dot{\theta}_{1} = \omega + \varepsilon \inner{\BZ(\theta_{1})}{\left( \Bf_{1}(\BX_{0}(\theta_{1})) + \BH_{1}(\theta_{1},\theta_{2}) \right)}, \\
\dot{\theta}_{2} = \omega + \varepsilon \inner{\BZ(\theta_{2})}{\left( \Bf_{2}(\BX_{0}(\theta_{2})) + \BH_{2}(\theta_{2},\theta_{1}) \right)}.
\end{aligned}
\end{align}
We remark that the functional differential equations~\eqref{eq:model} with a general coupling function depending on the past time series of the oscillators can be expressed as simple ordinary differential equations depending only on the current phases $\theta_{1,2}$ by use of the phase-reduction approximation.
Therefore, we can use the reduced phase equations, which are simply ordinary differential functions, for analyzing practical coupling schemes such as coupling with a time delay or coupling via temporal filters.


\subsection{Mutual synchronization}
\label{sec:model_3}

Once we derive the phase equations, we can analyze mutual synchronization of the two oscillators 
by the standard methods~\cite{Kuramoto1984chemical,Hoppensteadt1997weakly,Nakao2016phase}.
By introducing relative phase variables $\phi_{1,2}(t) = \theta_{1,2}(t) - \omega t \in \mathbb{R}$,
which are slowly varying because their derivatives $\dot{\phi}_{1,2}$ are of $O(\varepsilon)$,
we can conduct the averaging approximation of the coupling terms over one period of oscillation and obtain 
\begin{align}
\begin{aligned}
\dot{\phi}_{1} = \varepsilon \left( \Delta_{1} + \Gamma_{1}(\phi_{1} - \phi_{2}) \right), \\
\dot{\phi}_{2} = \varepsilon \left( \Delta_{2} + \Gamma_{2}(\phi_{2} - \phi_{1}) \right),
\end{aligned}
\end{align}
which are correct up to $O(\varepsilon)$.
Here, $\Gamma_{1,2}$ are the phase coupling functions (PCFs), which are $2\pi$-periodic functions defined by 
\begin{align}
\begin{aligned}
\Gamma_{1,2}(\varphi) &= \average{\inner{\BZ(\varphi + \psi)}{\BH_{1,2}(\varphi + \psi,\psi)}}{\psi}\\
&= \average{\inner{\BZ(\psi)}{\BH_{1,2}(\psi,\psi - \varphi)}}{\psi}.
\end{aligned}
\end{align}
Also, $\Delta_{1,2}$ are the deviations of the frequencies from the frequency $\omega$ of the common part $\BF$, represented as 
\begin{align}
\begin{aligned}
\Delta_{1,2} &= \average{\inner{\BZ(\psi)}{\Bf_{1,2}(\BX_{0}(\psi))}}{\psi}.
\end{aligned}
\end{align}
Here, we denote the averaging of a smooth function $g(\psi)$ over one period of oscillation as
\begin{align}
\average{g(\psi)}{\psi} = \frac{1}{2\pi}\int_{0}^{2\pi} g(\psi)d\psi.
\end{align}

To analyze mutual synchronization of the oscillators, we introduce the phase difference $\varphi = \phi_{1} - \phi_{2} \in \mathbb{R}$, which obeys
\begin{align}
\label{eq:phase_diff}
\dot{\varphi} = \varepsilon \left( \Delta + \Gamma_{\mathrm{d}}(\varphi) \right).
\end{align}
Here, $\Delta = \Delta_{1} - \Delta_{2}$ is a frequency mismatch between the oscillators, where we can assume $\Delta \geq 0$ without loss of generality,
and $\Gamma_{\mathrm{d}}(\varphi)$ is the difference between the PCFs $\Gamma_{1,2}$~(denoted hereafter as DPCF) defined as
\begin{align}
\begin{aligned}
\Gamma_{\mathrm{d}}(\varphi) &= \Gamma_{1}(\varphi) - \Gamma_{2}(-\varphi) \\
&= \average{\inner{\BZ(\psi)}{(\BH_{1}(\psi,\psi - \varphi) - \BH_{2}(\psi,\psi + \varphi))}}{\psi}.
\end{aligned}
\end{align}

If $\dot{\varphi} = 0$, the two oscillators are synchronized.
A stable phase-locking point $\varphi^{*}$ satisfies the following phase-locking condition~\eqref{eq:phase_locking} and stability condition~\eqref{eq:stability}:
\begin{align}
\label{eq:phase_locking}
\Delta + \Gamma_{\mathrm{d}}(\varphi^{*}) = 0, \\
\label{eq:stability}
\Gamma_{\mathrm{d}}'(\varphi^{*}) < 0,
\end{align}
where $\Gamma_{\mathrm{d}}'(\varphi)$ is the derivative of $\Gamma_{\mathrm{d}}(\varphi)$.
If a stable $\varphi^{*}$ exists, $\Gamma_{\mathrm{d}}$ also has at least one unstable fixed point $\overline{\varphi}$ within $[0,2\pi)$ from the continuity and periodicity of $\Gamma_{\mathrm{d}}$.

If both MCFs are identical, namely, $\BH_{1,2} = \BH$, we obtain $\Gamma_{1} = \Gamma_{2}$,
and $\Gamma_{\mathrm{d}}(\varphi)$ becomes twice the antisymmetric part of $\Gamma_{1,2}$ and is also a $2\pi$-periodic function.
We denote this $\Gamma_{\mathrm{d}}(\varphi)$ by $\Gamma_{\mathrm{a}}(\varphi)$, which can be represented as
\begin{align}
\begin{aligned}
\Gamma_{\mathrm{a}}(\varphi) 
&= \average{\inner{\BZ(\psi)}{(\BH(\psi,\psi - \varphi) - \BH(\psi,\psi + \varphi))}}{\psi}.
\end{aligned}
\label{eq:gamma_a}
\end{align}
We call this $\Gamma_{\mathrm{a}}$ (twice) the antisymmetric part of the PCF (APCF).
Moreover, if the two oscillators are identical, namely, $\BF_{1,2} = \BF$, we obtain $\Delta = 0$, so the phase difference obeys 
\begin{align}
\label{eq:phase_diff_sym}
\dot{\varphi} = \varepsilon \Gamma_{\mathrm{a}}(\varphi).
\end{align}
In this case, a stable phase-locking point $\varphi^{*}$ satisfies the following phase-locking condition~\eqref{eq:phase_locking_apcf} and stability condition~\eqref{eq:stability_apcf}:
\begin{align}
\label{eq:phase_locking_apcf}
\Gamma_{\mathrm{a}}(\varphi^{*}) = 0, \\
\label{eq:stability_apcf}
\Gamma'_{\mathrm{a}}(\varphi^{*}) < 0.
\end{align}
From the antisymmetry and $2\pi$-periodicity of $\Gamma_{\mathrm{a}}$, the phase-locking condition
$\Gamma_{\mathrm{a}}(\varphi) = 0$ is always satisfied in the cases of in-phase synchronization ($\varphi = 0$) and anti-phase synchronization ($\varphi = \pi$) within $\varphi \in [0,2\pi)$.


\section{Derivation of optimal coupling}
\label{sec:methods}

In this section, we propose a general method to optimize the MCFs that exhibit fast and global phase synchronization. 
We show that the optimal MCFs are proportional to the PSF.


\subsection{Symmetrically coupled identical oscillators}
\label{sec:methods_1}

\subsubsection{Representation of the mutual coupling function}
\label{sec:methods_1_1}

We first consider the case where identical oscillators are symmetrically coupled and assume that $\BF_{1,2} = \BF$ and $\tilde{\BH}_{1,2} = \tilde{\BH}$ (i.e., $\BH_{1,2} = \BH$).
In this case, the frequency mismatch is $\Delta = 0$, 
so the dynamics of the phase difference $\varphi$ is determined solely by the APCF $\Gamma_{\mathrm{a}}$ in Eq.~\eqref{eq:gamma_a}
and both in-phase and anti-phase synchronized solution always exist.
We need to consider $\Gamma_{\mathrm{a}}(\varphi)$ only on the interval $\varphi \in [-\pi,0]$;
$\Gamma_{\mathrm{a}}(\varphi)$ for $\varphi \in [0,\pi]$ is then determined from the antisymmetry.
We express the APCF as 
\begin{align}
\begin{aligned}
\label{eq:Gamma_H_pm}
\Gamma_{\mathrm{a}}(\varphi) &= \average{\inner{\BZ(\psi)}{(\BH(\psi,\psi - \varphi) - \BH(\psi,\psi + \varphi))}}{\psi} \\
&= \average{\inner{\BZ(\psi)}{(\BH^{-\varphi}(\psi) - \BH^{+\varphi}(\psi))}}{\psi}
\end{aligned}
\end{align}
by introducing the following representations of the MCF $\BH$:
\begin{align}
\begin{aligned}
\BH^{+\varphi}(\psi) &= \BH(\psi,\psi + \varphi), \\
\BH^{-\varphi}(\psi) &= \BH(\psi,\psi - \varphi),
\end{aligned}
\end{align}
which are both $2\pi$-periodic in $\varphi$ and $\psi$.
We can independently optimize $\Gamma_{\mathrm{a}}(\varphi)$ at each value of $\varphi \in [-\pi,0]$, 
because $\BH^{+\varphi}(\psi)$ and $\BH^{+\tilde{\varphi}}(\psi)$ have no overlap with each other when $\varphi \neq \tilde{\varphi}$ for $\varphi, \tilde{\varphi} \in [-\pi,0]$.
The same holds true for $\BH^{-\varphi}(\psi)$.

\subsubsection{Optimal functional form of the mutual coupling function}
\label{sec:methods_1_2}

We first consider stable in-phase synchronization and seek an energy-efficient functional form of the MCF.
To realize the fastest convergence of the phase difference $\varphi$ towards the stable in-phase synchronized state $\varphi^{*} = 0$ at each value of $\varphi$, 
we maximize the APCF $\Gamma_{\mathrm{a}}(\varphi)$ under a given average input power $P(\varphi)^{2}$ of the functions $\BH^{\pm \varphi}(\psi)$, where $P(\varphi) \geq 0$.
From the antisymmetry of $\Gamma_{\mathrm{a}}(\varphi)$, this yields the fastest convergence towards $\varphi^{*} = 0$ also for $\varphi \in [0, \pi]$.

This optimization problem at each fixed value of $\varphi \in [-\pi,0]$ is formulated as follows:
\begin{align}
\begin{aligned}
&\max_{\BH^{+\varphi},\BH^{-\varphi}} && \Gamma_{\mathrm{a}}(\varphi) \\
&\mathrm{s.t.} && \frac{1}{2} \average{\norm{\BH^{+\varphi}(\psi)}^{2} + \norm{\BH^{-\varphi}(\psi)}^{2}}{\psi} = P(\varphi)^{2},
\end{aligned}
\end{align}
where the norm is defined as $\norm{\bm{a}} = \sqrt{\sum_{i=1}^{N} a_{i}^{2}}$.
This optimization problem can be analytically solved by the method of Lagrange multipliers.
We introduce a functional
\begin{align}
\begin{aligned}
&S\left\{ \BH^{+\varphi},\BH^{-\varphi},\lambda \right\} \\
={}& \average{\inner{\BZ(\psi)}{(\BH^{-\varphi}(\psi) - \BH^{+\varphi}(\psi))}}{\psi} \\
&+ \lambda \left( \frac{1}{2} \average{\norm{\BH^{+\varphi}(\psi)}^{2} + \norm{\BH^{-\varphi}(\psi)}^{2}}{\psi} - P(\varphi)^{2} \right),
\end{aligned}
\end{align}
where $\lambda$ is a Lagrange multiplier.
From the extremum conditions for $S$ with respect to $\BH^{\pm \varphi}$:
\begin{align}
\begin{aligned}
\pd{S}{\BH^{+\varphi}} &= -\BZ(\psi) + \lambda \BH^{+\varphi}(\psi) = 0, \\
\pd{S}{\BH^{-\varphi}} &= \BZ(\psi) + \lambda \BH^{-\varphi}(\psi) = 0,
\end{aligned}
\end{align}
we obtain 
\begin{align}
\begin{aligned}
\BH^{+\varphi}(\psi) &= \frac{1}{\lambda}\BZ(\psi), \\
\BH^{-\varphi}(\psi) &= -\frac{1}{\lambda}\BZ(\psi),
\end{aligned}
\end{align}
where we find that $\BH^{+\varphi}$ and $\BH^{-\varphi}$ are proportional to the PSF $\BZ$.
Considering the constraint on the input power, the Lagrange multiplier should satisfy 
\begin{align}
\lambda^{2} = \frac{\average{\norm{\BZ(\psi)}^{2}}{\psi}}{P(\varphi)^{2}}.
\end{align}

Thus, the optimal functional forms of $\BH^{\pm \varphi}$ are obtained as 
\begin{align}
\label{eq:opt_Hpm}
\BH^{+\varphi}(\psi) = -\BH^{-\varphi}(\psi) = -\frac{P(\varphi) \BZ(\psi)}{\sqrt{\average{\norm{\BZ(\psi)}^{2}}{\psi}}},
\end{align}
where the negative value of $\lambda$ should be chosen for maximizing $\Gamma_{\mathrm{a}}(\varphi)$.
We note that the optimal MCF in Eq.~\eqref{eq:opt_Hpm} is proportional to $\BZ(\psi)$,
namely, it is given as a product of $P(\varphi)$ and the waveform of the PSF $\BZ(\psi)$ for each $\varphi \in [-\pi, 0]$. 
In this sense, we call $P(\varphi)$ the amplitude of the MCF in what follows. 
In the original variables, the optimal MCF can be represented as 
\begin{align}
\label{eq:H12}
\BH(\theta_{1},\theta_{2}) = \frac{\BZ(\theta_{1}) P(\theta_{1} - \theta_{2})}{\sqrt{\average{\norm{\BZ(\psi)}^{2}}{\psi}}}.
\end{align}
From Eq.~\eqref{eq:Gamma_H_pm}, the optimal APCF $\Gamma_{\mathrm{a}}(\varphi)$ with the above $\BH^{\pm \varphi}(\psi)$ can be represented as 
\begin{align}
\label{eq:opt_Gamma_a}
\Gamma_{\mathrm{a}}(\varphi) = 2P(\varphi)\sqrt{\average{\norm{\BZ(\psi)}^{2}}{\psi}} = CP(\varphi),
\end{align}
where $C = 2\sqrt{\average{\norm{\BZ(\psi)}^{2}}{\psi}}$.

It is remarkable that the optimized $\Gamma_{\mathrm{a}}(\varphi)$ is simply proportional to the amplitude $P(\varphi)$ for each $\varphi$ and depends only on the averaged norm $\average{\norm{\BZ(\psi)}^{2}}{\psi}$ over one period, not on the functional form, of the PSF $\BZ(\psi)$.
Therefore, we only need to obtain $P(\varphi)$ and use it in Eq.~\eqref{eq:opt_Gamma_a} for determining the full functional form of $\Gamma_{\mathrm{a}}(\varphi)$ for $\varphi \in [-\pi,0]$.
Thus, because the MCF is factorized into the oscillator-specific part (PSF) and the oscillator-independent part (amplitude), we can realize fast and global synchronization irrespective of the detailed characteristics of the limit-cycle oscillators by finding the optimal amplitude $P(\varphi)$.

\subsubsection{Optimization of the amplitude}
\label{sec:methods_1_3}

As explained above, the optimal MCF that yields the fastest convergence at each $\varphi$ is simply proportional to $\BZ(\psi)$ and we only need to determine the amplitude $P(\varphi)$. 
We now optimize the amplitude $P(\varphi)$ for $\varphi \in [-\pi,0]$ to minimize the average convergence time to the stable in-phase synchronized state.
Since $\Gamma_{\mathrm{a}}(\varphi)$ can be independently chosen at each value of $\varphi \in [-\pi,0]$,
we can assume that $\Gamma_{\mathrm{a}}(\varphi) \neq 0$ for all $\varphi \in (-\pi,0)$ (note that $\Gamma_{\mathrm{a}}(-\pi) = \Gamma_{\mathrm{a}}(0) = 0$).

Therefore, as in~\cite{Kato2021optimization}, which addressed optimization of periodic driving signals for fast entrainment of a single oscillator, 
we can define the average convergence time $T_\mathrm{ave}$ of the phase difference between the two oscillators by
\begin{align}
\begin{aligned}
T_{\mathrm{ave}} &= \frac{1}{\pi}\int_{-\pi + \varepsilon_{\overline{\varphi}}}^{-\varepsilon_{\varphi^{*}}} \int_{\tilde{\varphi}}^{-\varepsilon_{\varphi^{*}}} \frac{1}{\dot{\varphi}} d\varphi d\tilde{\varphi} \\
&= \frac{1}{\pi \varepsilon}\int_{-\pi + \varepsilon_{\overline{\varphi}}}^{-\varepsilon_{\varphi^{*}}} \int_{\tilde{\varphi}}^{-\varepsilon_{\varphi^{*}}} \frac{1}{\Gamma_{\mathrm{a}}(\varphi)} d\varphi d\tilde{\varphi} \\
&= \frac{1}{\pi \varepsilon C}\int_{-\pi + \varepsilon_{\overline{\varphi}}}^{-\varepsilon_{\varphi^{*}}} \int_{\tilde{\varphi}}^{-\varepsilon_{\varphi^{*}}} \frac{1}{P(\varphi)} d\varphi d\tilde{\varphi},
\end{aligned}
\end{align}
where the first integral $ (1/\pi)\int_{-\pi + \varepsilon_{\overline{\varphi}}}^{-\varepsilon_{\varphi^{*}}} d\tilde{\varphi}$ represents an average over the initial phase difference $\tilde\varphi$ uniformly distributed in the interval $[-\pi + \varepsilon_{\overline{\varphi}},\; -\varepsilon_{\varphi^{*}}]$, 
and the second integral $\int_{\tilde{\varphi}}^{-\varepsilon_{\varphi^{*}}} d\varphi$ of $(1/\dot{\varphi})$ represents the time necessary for $\varphi$ 
to converge from the initial phase difference $\tilde{\varphi}$ to a small acceptance range $[ -\varepsilon_{\varphi^{*}}, 0 ]$ near the in-phase synchronized state from the negative side.
Here, $\varepsilon_{\varphi^{*}}$ and $\varepsilon_{\overline{\varphi}}$ ($0 < \varepsilon_{\varphi^{*}},\; \varepsilon_{\overline{\varphi}} \ll 1$) are small parameters for preventing the divergence of the integral at the stable fixed point ($\varphi^{*} = 0$) and unstable fixed point ($\overline{\varphi} = -\pi$), respectively.
This is because the time necessary for the phase difference $\varphi$ to reach the stable fixed point or escape from the unstable fixed point diverges for smooth $\Gamma_{\mathrm{a}}$.
Thus, $\varepsilon_{\varphi^{*}}$ is the threshold value below which we judge that the two oscillators are in-phase synchronized and, similarly, $\varepsilon_{\overline{\varphi}}$ is introduced to exclude the initial conditions near the unstable fixed point.
The average convergence time $T_{\mathrm{ave}}$ for $\varphi \in [0,\pi]$ is the same as that for $\varphi \in [-\pi,0]$ because of the antisymmetry of the APCF.

We can formulate the optimization problem to minimize the average convergence time $T_{\mathrm{ave}}$ towards $\varphi^{*} = 0$ under a given total average input power $Q$ as follows:
\begin{align}
\begin{aligned}
\label{prob:opt_P}
&\min_{P} && \int_{-\pi + \varepsilon_{\overline{\varphi}}}^{-\varepsilon_{\varphi^{*}}} \int_{\tilde{\varphi}}^{-\varepsilon_{\varphi^{*}}} \frac{1}{P(\varphi)} d\varphi d\tilde{\varphi}
+ \gamma \int_{-\pi}^{0} \left( \frac{dP(\varphi)}{d\varphi} \right)^{2} d\varphi \\
&\mathrm{s.t.} && \frac{1}{\pi}\int_{-\pi}^{0} P(\varphi)^{2} d\varphi = Q, \\
&&& P(\varphi) \geq 0 \quad (\varphi \in [-\pi,0]), \\
&&& P(0) = 0, \\
&&& P(-\pi) = 0,
\end{aligned}
\end{align}
where we introduced the integral of the squared gradient of $P(\varphi)$ over $\varphi \in [-\pi,0]$ to the objective function with a weight $\gamma > 0$ in order to suppress sharp variations. 
As shown in Appendix B, this regularization term suppresses the discontinuity of $P(\varphi)$ near $\varphi^{*} = 0$. 
The third constraint represents the condition for the stable fixed point and the fourth constraint represents that for the unstable fixed point, i.e., $\Gamma_{\mathrm{a}}(0) = C P(0) = 0$ and $\Gamma_{\mathrm{a}}(-\pi) = C P(-\pi) = 0$ from Eq.~\eqref{eq:opt_Gamma_a}, respectively.

It is difficult to solve this problem analytically, so we numerically obtain $P(\varphi)$ by discretizing its functional form as $P(\varphi_{m}) = P_{m}\; (m = 0,1,\dots,M)$, where $\varphi \in [-\pi,0]$ is discretized as $\varphi_{m} = -m\Delta_{\varphi}\; (m = 0,1,\dots,M)$ with an interval of $\Delta_{\varphi} = \pi / M$.
The numerical optimization problem is then
\begin{align}
\begin{aligned}
\label{prob:opt_P_num}
&\min_{\{P_{m}\}} && \Delta_{\varphi}^{2} \sum_{m=1}^{M-1} \sum_{j=1}^{m} \frac{1}{P_{j}} + \gamma \frac{1}{\Delta_{\varphi}} \sum_{m=1}^{M} (P_{m} - P_{m-1})^{2} \\
&\mathrm{s.t.} && \frac{1}{M}\sum_{m=0}^{M} P_{m}^{2} = Q, \\
&&& P_{m} \geq 0 \quad (m = 0,1,\dots,M), \\
&&& P_{0} = 0, \\
&&& P_{M} = 0.
\end{aligned}
\end{align}
We denote the optimal $P_{m}$ obtained numerically by $P_{\mathrm{opt}}(\varphi_{m};\gamma)$.
The optimal functional form of $P_{\mathrm{opt}}(\varphi_{m};\gamma)$ for $\varphi_{m} \in [0,\pi]$ is given by $-P_{\mathrm{opt}}(-\varphi_{m};\gamma)$ for $\varphi_{m} \in [-\pi,0]$ from the antisymmetry of the APCF.
Using the obtained $P_{\mathrm{opt}}(\varphi_{m};\gamma)$, we can calculate $\BH^{\pm \varphi}(\psi)$ by Eq.~\eqref{eq:opt_Hpm} and $\Gamma_{\mathrm{a}}(\varphi_{m})$ by Eq.~\eqref{eq:opt_Gamma_a} at each $\varphi_{m}$, respectively. 

The above formulation aims to achieve stable in-phase synchronization ($\varphi^{*} = 0$).
If we consider realizing stable anti-phase synchronization ($\varphi^{*} = \pi$) instead, we can obtain the optimal solution 
as $P_{\mathrm{opt}}(\varphi + \pi;\gamma)$ by shifting the optimal solution $P_{\mathrm{opt}}(\varphi;\gamma)$ for stable in-phase synchronization by $\pi$.


\subsection{Nearly identical oscillators}
\label{sec:methods_2}

\subsubsection{Representation of the mutual coupling function}
\label{sec:methods_2_1}

In this subsection, we generalize the method for a pair of oscillators with nearly identical (slightly nonidentical) dynamics $\BF_{1,2}$ interacting through different $\tilde{\BH}_{1,2}$, 
while we considered a pair of identical oscillators with symmetric coupling ($\BF_{1,2} = \BF$, $\tilde{\BH}_{1,2} = \BH$) in the previous subsection. 

Since $\Gamma_{\mathrm{d}}(\varphi)$ is $2\pi$-periodic but no longer antisymmetric in general,
we need to consider $\Gamma_{\mathrm{d}}(\varphi)$ on the whole interval $\varphi \in [0,2\pi)$.
We express the DPCF as
\begin{align}
\begin{aligned}
\Gamma_{\mathrm{d}}(\varphi) &= \average{\inner{\BZ(\psi)}{(\BH_{1}(\psi,\psi - \varphi) - \BH_{2}(\psi,\psi + \varphi))}}{\psi} \\
&= \average{\inner{\BZ(\psi)}{(\BH_{1}^{-\varphi}(\psi) - \BH_{2}^{+\varphi}(\psi))}}{\psi}
\end{aligned}
\end{align}
by introducing the following representations of the MCFs $\BH_{1,2}$:
\begin{align}
\begin{aligned}
\BH_{1}^{-\varphi}(\psi) &= \BH_{1}(\psi,\psi - \varphi), \\
\BH_{2}^{+\varphi}(\psi) &= \BH_{2}(\psi,\psi + \varphi).
\end{aligned}
\end{align}
We can optimize $\Gamma_{\mathrm{d}}(\varphi)$ at each value of $\varphi \in [0,2\pi)$ independently, 
because $\BH_{1}^{-\varphi}(\psi)$ and $\BH_{1}^{-\tilde{\varphi}}(\psi)$ have no overlap with each other when $\varphi \neq \tilde{\varphi}$ for $\varphi, \tilde{\varphi} \in [0,2\pi)$.
This is also the case for $\BH_{2}^{+\varphi}(\psi)$.

\subsubsection{Optimal functional form of the mutual coupling function}
\label{sec:methods_2_2}

Without loss of generality, we can assume the target phase difference to be zero, i.e., in-phase synchronization, 
because $\Gamma_{\mathrm{d}}(\varphi)$ can be chosen independently at each $\varphi$.
That is, if the target phase difference is originally $\varphi^{*}$, we optimize the DPCF $\Gamma_{\mathrm{d}}(\varphi)$ for stable in-phase synchronization and then shift it as $\Gamma_{\mathrm{d}}(\varphi - \varphi^{*})$.
Note that we are considering $\Gamma_{\mathrm{d}}(\varphi)$ on the whole interval $\varphi \in [0, 2\pi)$. 
Thus, it is no longer antisymmetric and can take both positive and negative values, in contrast to the previous case for identical oscillators where $\Gamma_{\mathrm{a}}(\varphi)$ was considered only for $\varphi \in [-\pi, 0]$ and assumed always positive. 

As in the previous case, at each fixed $\varphi$, we maximize the absolute value of the DPCF $| \Gamma_{\mathrm{d}}(\varphi) |$ under a given average input power $P(\varphi)^{2}$ of the MCFs $\BH_{1}^{-\varphi}$ and $\BH_{2}^{+\varphi}$. 
Here, because $\Gamma_{\mathrm{d}}(\varphi)$ can take both positive and negative values, the absolute value of $\Gamma_{\mathrm{d}}(\varphi)$ is maximized to make efficient use of the given input power.
This problem is formulated as 
\begin{align}
\begin{aligned}
&\max_{\BH_{1}^{-\varphi},\BH_{2}^{+\varphi}} && |\Gamma_{\mathrm{d}}(\varphi)| \\
&\mathrm{s.t.} && \frac{1}{2} \average{\norm{\BH_{1}^{-\varphi}(\psi)}^{2} + \norm{\BH_{2}^{+\varphi}(\psi)}^{2}}{\psi} = P(\varphi)^{2}.
\end{aligned}
\end{align}
It is easy to see that we need to maximize $\Gamma_{\mathrm{d}}(\varphi)$ when $P(\varphi) \geq 0$ and minimize $\Gamma_{\mathrm{d}}(\varphi)$ when $P(\varphi) \leq 0$.
As in the previous case, we find that the optimal functional forms of the MCFs are given in the form
\begin{align}
\BH_{1}^{-\varphi}(\psi) = -\BH_{2}^{+\varphi}(\psi) = \frac{P(\varphi)\BZ(\psi)}{\sqrt{\average{\norm{\BZ(\psi)}^{2}}{\psi}}},
\end{align}
which can also be represented in the original variables as 
\begin{align}
\BH_{1}(\theta_{1},\theta_{2}) &= \frac{\BZ(\theta_{1}) P(\theta_{1} - \theta_{2})}{\sqrt{\average{\norm{\BZ(\psi)}^{2}}{\psi}}}, \\
\BH_{2}(\theta_{1},\theta_{2}) &= -\frac{\BZ(\theta_{1}) P(\theta_{2} - \theta_{1})}{\sqrt{\average{\norm{\BZ(\psi)}^{2}}{\psi}}}.
\end{align}
The optimal DPCF is obtained as 
\begin{align}
\label{eq:Gamma_prop_P}
\Gamma_{\mathrm{d}}(\varphi) = 2P(\varphi)\sqrt{\average{\norm{\BZ(\psi)}^{2}}{\psi}} = CP(\varphi),
\end{align}
where we can again find that the optimal DPCF $\Gamma_{\mathrm{d}}(\varphi)$ is proportional to $P(\varphi)$ and
does not depend on the functional form of $\BZ(\psi)$.

\subsubsection{Optimization of the amplitude}
\label{sec:methods_2_3}

We next optimize $P(\varphi)$ by minimizing the average convergence time.
Since the value of $P(\varphi)$ can be independently chosen at each $\varphi \in [0,2\pi)$,
we can assume that only one unstable fixed point $\overline{\varphi}$ of Eq.~\eqref{eq:phase_diff} exists in $(0, 2\pi)$.
The average convergence time $T_{\mathrm{ave}}$ is then defined as
\begin{align}
\begin{aligned}
&T_{\mathrm{ave}} \\
={}& \frac{1}{2\pi} \left( \int_{\overline{\varphi} - 2\pi + \varepsilon_{\overline{\varphi}}}^{-\varepsilon_{\varphi^{*}}} \int_{\tilde{\varphi}}^{-\varepsilon_{\varphi^{*}}} \frac{1}{\dot{\varphi}} d\varphi d\tilde{\varphi} \right. 
+ \left. \int_{\varepsilon_{\varphi^{*}}}^{\overline{\varphi} - \varepsilon_{\overline{\varphi}}} \int_{\tilde{\varphi}}^{\varepsilon_{\varphi^{*}}} \frac{1}{\dot{\varphi}} d\varphi d\tilde{\varphi} \right) \\
={}& \frac{1}{2\pi\varepsilon} \left( \int_{\overline{\varphi} - 2\pi + \varepsilon_{\overline{\varphi}}}^{-\varepsilon_{\varphi^{*}}} \int_{\tilde{\varphi}}^{-\varepsilon_{\varphi^{*}}} \frac{1}{\Delta + \Gamma_{\mathrm{d}}(\varphi)} d\varphi d\tilde{\varphi} \right. \\
&+ \left. \int_{\varepsilon_{\varphi^{*}}}^{\overline{\varphi} - \varepsilon_{\overline{\varphi}}} \int_{\tilde{\varphi}}^{\varepsilon_{\varphi^{*}}} \frac{1}{\Delta + \Gamma_{\mathrm{d}}(\varphi)} d\varphi d\tilde{\varphi} \right)
\end{aligned}
\end{align}
by summing up the convergence times of uniformly distributed initial phase differences on 
$[\overline{\varphi}-2\pi,\overline{\varphi}]$ (except for the regions near the fixed points) to the in-phase synchronization state (more precisely to $[-\varepsilon_{\varphi^{*}}, +\varepsilon_{\varphi^{*}}]$ around $\varphi^*=0$) from negative and positive sides,
where we again introduced the sufficiently small parameters $0 < \varepsilon_{\varphi^{*}}, \varepsilon_{\overline{\varphi}} \ll 1$ for preventing the divergence of the integrals near the stable and unstable fixed points.
In this case, $T_{\mathrm{ave}}$ consists of two terms, where the first term represents the sum of the convergence time of $\varphi$ from $[{\overline{\varphi} - 2\pi + \varepsilon_{\overline{\varphi}}}, {-\varepsilon_{\varphi^{*}}}]$ to $-\varepsilon_{\varphi^{*}}$, i.e., from the negative side of $\varphi^{*} = 0$, 
and the second term represents that from $[{\varepsilon_{\varphi^{*}}}, {\overline{\varphi} - \varepsilon_{\overline{\varphi}}}]$ to ${\varepsilon_{\varphi^{*}}}$, i.e., from the positive side of $\varphi^{*} = 0$. 
We note that we now consider the interval $[\overline{\varphi}-2\pi,\overline{\varphi}]$ instead of $[0,2\pi]$.

The minimization problem of $T_{\mathrm{ave}}$ under a given total average input power $Q$ is formulated as 
\begin{align}
\begin{aligned}
&\min_{\overline{\varphi},P} && \int_{\overline{\varphi} - 2\pi + \varepsilon_{\overline{\varphi}}}^{-\varepsilon_{\varphi^{*}}} \int_{\tilde{\varphi}}^{-\varepsilon_{\varphi^{*}}} \frac{1}{\Delta + \Gamma_{\mathrm{d}}(\varphi)} d\varphi d\tilde{\varphi} \\
&&& + \int_{\varepsilon_{\varphi^{*}}}^{\overline{\varphi} - \varepsilon_{\overline{\varphi}}} \int_{\tilde{\varphi}}^{\varepsilon_{\varphi^{*}}} \frac{1}{\Delta + \Gamma_{\mathrm{d}}(\varphi)} d\varphi d\tilde{\varphi} \\
&&& + \gamma \int_{\overline{\varphi} - 2\pi}^{\overline{\varphi}} \left( \frac{dP(\varphi)}{d\varphi} \right)^{2} d\varphi \\
&\mathrm{s.t.} && \frac{1}{2\pi}\int_{\overline{\varphi} - 2\pi}^{\overline{\varphi}} P(\varphi)^{2} d\varphi = Q, \\
&&& \Delta + \Gamma_{\mathrm{d}}(\varphi) \leq 0 \quad (\varphi \in [0,\overline{\varphi}]), \\
&&& \Delta + \Gamma_{\mathrm{d}}(\varphi) \geq 0 \quad (\varphi \in [\overline{\varphi} - 2\pi,0]), \\
&&& \Delta + \Gamma_{\mathrm{d}}(0) = 0, \\
&&& \Delta + \Gamma_{\mathrm{d}}(\overline{\varphi} - 2\pi) = 0, \\
&&& \Delta + \Gamma_{\mathrm{d}}(\overline{\varphi}) = 0.
\end{aligned}
\end{align}
Note that we optimize not only the amplitude $P(\varphi)$ of the MCFs but also the location of the unstable fixed point $\overline{\varphi}$ in the present case, as $\overline{\varphi}$ can be chosen arbitrarily and affects the average convergence time $T_{\mathrm{ave}}$.
As before, we introduced the integral of the squared gradient of $P(\varphi)$ on $\varphi \in [\overline{\varphi}-2\pi,\overline{\varphi}]$ to the objective function with a weight $\gamma > 0$ as the regularization term to avoid sharp variations of $P(\varphi)$, in particular, the discontinuity near $\varphi^{*} = 0$. 

Since this problem is also difficult to solve analytically, we numerically obtain $P(\varphi)$ by discretizing its functional form as $P(\varphi_{k}) = P_{k}\; (k = 0,1,\dots,K)$, 
where $\varphi \in [\overline{\varphi}-2\pi,\overline{\varphi}]$ is discretized as $\varphi_{m} = k\Delta_{\varphi} + \overline{\varphi} - 2\pi\; (k = 0,1,\dots,K)$ with the interval of $\Delta_{\varphi} = 2\pi / K$.
The numerical optimization problem is then
\begin{align}
\begin{aligned}
\label{prob:opt_P_k_num}
&\min_{k^{*},\{ P_{k} \}} && \Delta_{\varphi}^{2} \sum_{k=1}^{k^{*}-1} \sum_{j=1}^{k} \frac{1}{\Delta + CP_{j}} \\
&&& - \Delta_{\varphi}^{2} \sum_{k=k^{*}+1}^{K-1} \sum_{j=k^{*}+1}^{k} \frac{1}{\Delta + CP_{j}} \\
&&& + \gamma \frac{1}{\Delta_{\varphi}} \sum_{k=1}^{K} \left( P_{k} - P_{k-1} \right)^{2} \\
&\mathrm{s.t.} && \frac{1}{K} \sum_{k=0}^{K} P_{k}^{2} = Q, \\
&&& P_{k} \leq -\frac{\Delta}{C} \quad (k = k^{*},\dots,K), \\
&&& P_{k} \geq -\frac{\Delta}{C} \quad (k = 0,\dots,k^{*}), \\
&&& P_{k^{*}} = -\frac{\Delta}{C}, \\
&&& P_{0} = -\frac{\Delta}{C}, \\
&&& P_{K} = -\frac{\Delta}{C},
\end{aligned}
\end{align}
where the problem is now converted to the optimization of the location of the stable fixed point $k^{*}$ and the function $\{ P_{k} \}$.
Once we obtain the optimal $k^{*}$, we determine $\overline{\varphi}$ so that $\varphi_{k^{*}} = 0$ holds.

In practice, we used ternary search for finding the optimal $k^{*}$, which can efficiently find local minima of a given objective function without calculating derivatives.
See Appendix~C for the details of numerical optimization. 
We note that, if $|\Delta|$ is too large for a given total average input power $Q$,
there is no feasible solution because $\Delta + \Gamma_{\mathrm{d}}(\varphi) = 0$ cannot be realized for all $\varphi$.


\section{Results}
\label{sec:results}

In this section, we illustrate how the proposed optimization method works using two types of limit-cycle oscillators and compare the performance with the previous optimization methods.


\subsection{Methods for evaluation}
\label{sec:results_1}

We now demonstrate the results of the proposed optimization method for two types of limit-cycle oscillators and compare the results with those for the previous optimization methods in~\cite{Watanabe2019optimization} that assume drive-response structures of the MCF (see Appendix~A). 
We call the latter `drive-response methods' in what follows.
Before showing the numerical results, we explain two methods for evaluating the performance.
We optimize the average convergence time in the present study, while the linear stability of the in-phase synchronized state is optimized in~\cite{Watanabe2019optimization}.
Therefore, we evaluate both the linear stability (for the identical cases) and the average convergence time.
When we compare the proposed method to the previous drive-response methods, where we optimized either the response matrix or the driving function, the response matrices were rescaled so that $\average{\average{\norm{\BH(\psi_{1},\psi_{2})}^{2}}{\psi_{1}}}{\psi_{2}} = Q$ in both cases to make the conditions equal.

The linear stability is characterized by the slope of the APCF at the synchronized state $\varphi^{*}$.
Since the functional form of the optimal APCF $\Gamma_{\mathrm{a}}(\varphi)$ is only 
numerically obtained in the present method, the value of the linear stability $\Gamma_{\mathrm{a}}'(\varphi)$ cannot be obtained analytically.
We therefore use 
\begin{align}
\tilde{\Gamma}'_{\mathrm{a}}(\varphi^{*}) = \frac{\Gamma_{\mathrm{a}}(\varphi^{*} + \Delta_{\varphi}) - \Gamma_{\mathrm{a}}(\varphi^{*} - \Delta_{\varphi})}{2\Delta_{\varphi}}
\end{align}
as the approximate value of the linear stability, where $\Delta_{\varphi}$ is the discretization interval introduced in Sec.~\ref{sec:methods}.
For the previous methods, we can calculate the linear stability analytically,
but we evaluate the value of the linear stability by the above equation for a fair comparison.

The convergence time to the synchronized state is measured by excluding the small regions near the fixed points.
Since the system state of smooth one-dimensional dynamics does not generally converge to a fixed point in a finite time, 
we regard the minimum time to reach the interval $[\varphi^{*} - \varepsilon_{\varphi^{*}},\varphi^{*} + \varepsilon_{\varphi^{*}}]$ that is sufficiently close to the synchronized state as the convergence time.
The convergence time is averaged over different initial conditions.


\subsection{FitzHugh-Nagumo oscillator}
\label{sec:results_2}

\subsubsection{Symmetrically coupled identical oscillators}
\label{sec:results_2_1}

\begin{figure}
\centering
\includegraphics[width=0.48\textwidth]{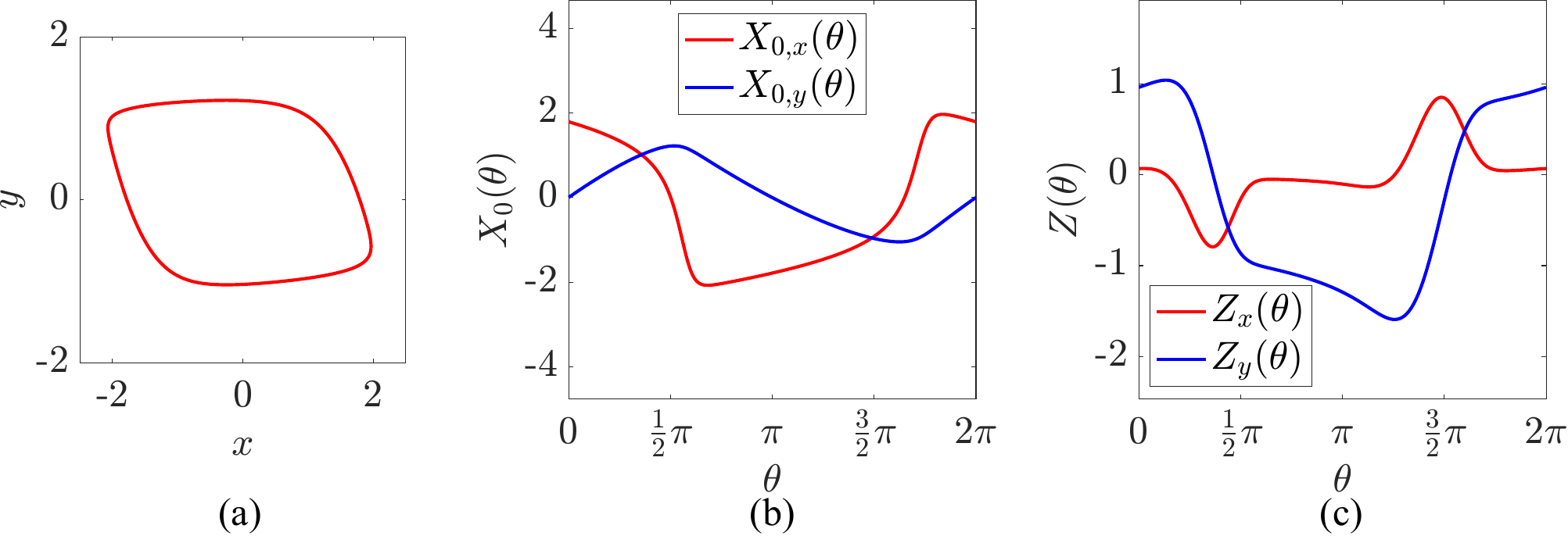}
\caption{
Limit cycle and PSF of the FHN oscillator.
(a)~Limit cycle on the $xy$ plane.
(b)~Limit-cycle solution $\BX = [X_{0,x}\; X_{0,y}]^{\top}$ vs. phase $\theta$.
(c)~PSF $\BZ = [Z_{x}\; Z_{y}]^{\top}$ vs. phase $\theta$.
}
\label{fig:fig1}
\end{figure}

We first consider a pair of weakly and symmetrically coupled FitzHugh-Nagumo (FHN) oscillators~\cite{fitzhugh1961impulses,Nagumo1962Active} with identical properties, given by Eq.~\eqref{eq:model}, where $\BF_{1,2} = \BF$ (i.e., $\Bf_{1,2} = \bm{0}$) and $\tilde{\BH}_{1,2} = \tilde{\BH}$ (i.e., $\BH_{1,2} = \BH$).
The FHN oscillator is represented as 
\begin{align}
\dot{\BX} = 
\begin{bmatrix}
\dot{x} \\ \dot{y}
\end{bmatrix}
= 
\begin{bmatrix}
x - a x^{3} - y \\
(x + b) c
\end{bmatrix}
,
\end{align}
where we assume $(a, b, c) = (1/3, 0.25, 0.15)$.
This oscillator has a limit cycle with a period $T = 21.94$ and natural frequency $\omega = 0.2864$.
We show the limit cycle on the $xy$ plane in Fig.~\ref{fig:fig1}(a), limit-cycle solution as a function of the phase in Fig.~\ref{fig:fig1}(b), and PSF calculated by the adjoint equation in Fig.~\ref{fig:fig1}(c), respectively.

\begin{figure}
\centering
\includegraphics[width=0.3\textwidth]{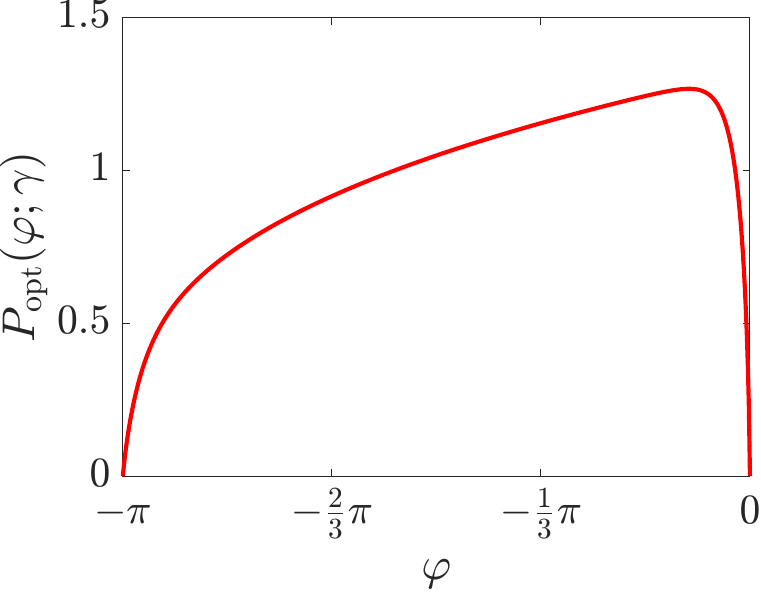}
\caption{Functional form of the optimal amplitude $P_{\mathrm{opt}}(\varphi;\gamma)$ for the coupled FHN oscillators ($Q = 1$, $M = 600$, and $\gamma = 10^{-2}$).}
\label{fig:fig2}
\end{figure}

\begin{figure}
\centering
\includegraphics[width=0.48\textwidth]{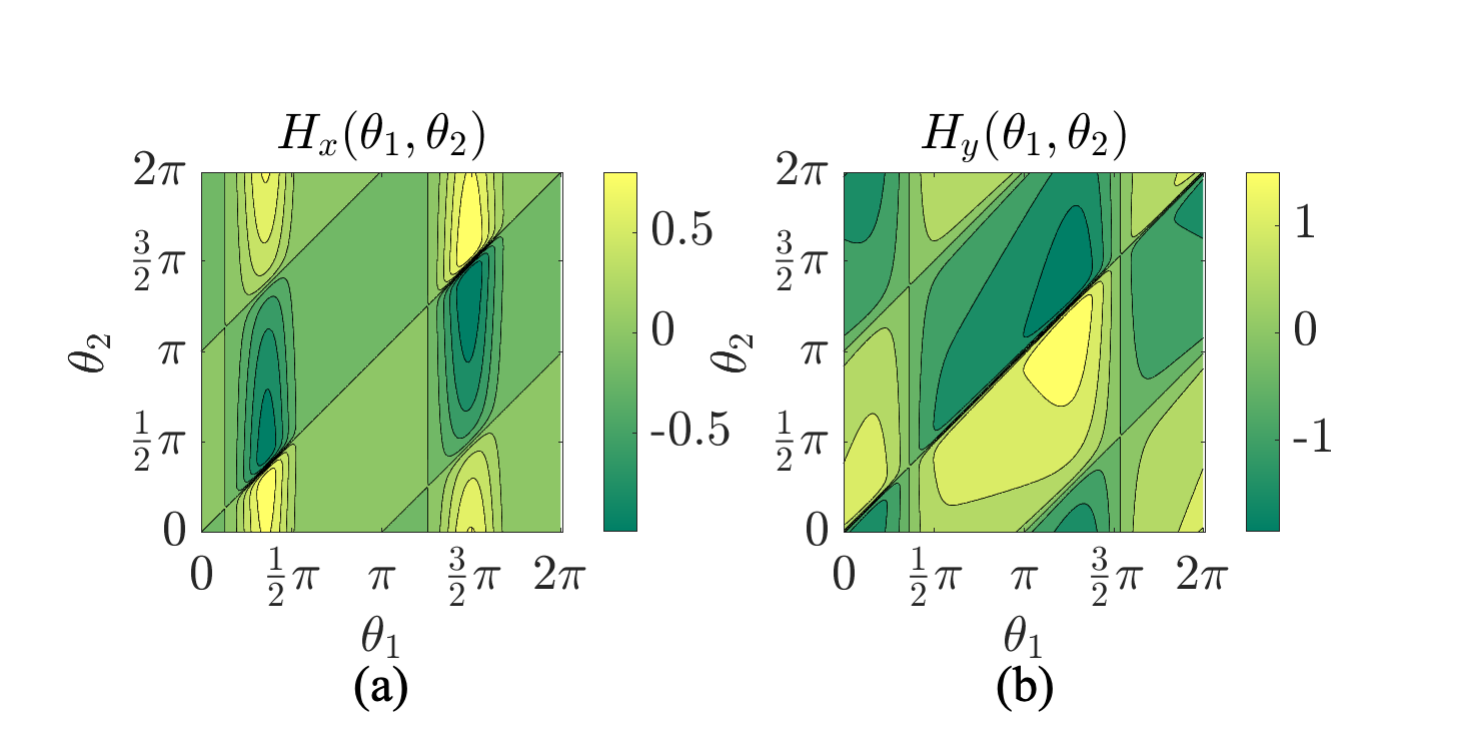}
\caption{
Optimal MCF $\BH(\theta_{1},\theta_{2}) = \left[ H_{x}(\theta_{1},\theta_{2})\; H_{y}(\theta_{1},\theta_{2}) \right]^{\top}$ by the proposed method for the symmetrically coupled identical FHN oscillators.
(a)~$x$ component $H_x$.
(b)~$y$ component $H_y$.
In each figure, the straight lines represent $\BH_{x,y}(\theta_{1},\theta_{2}) = 0$.
}
\label{fig:fig3}
\end{figure}

We first show the results by the proposed method.
Figure~\ref{fig:fig2} shows the optimal amplitude $P_{\mathrm{opt}}(\varphi;\gamma)$, where $Q = 1$, $M = 600$, and $\gamma = 10^{-2}$.
The small parameters in the objective functions are chosen as $\varepsilon_{\overline{\varphi}} = \varepsilon_{\varphi^{*}} = \pi/M$. 
The optimal amplitude $P_{\mathrm{opt}}(\varphi;\gamma)$ takes $0$ at $\varphi = -\pi$, increases with $\varphi$, and then suddenly drops to $0$ near $\varphi=0$. We can confirm that $P_{\mathrm{opt}}(\varphi;\gamma)$ does not exhibit discontinuity near $\varphi = 0$ owing to the additional regularization term in Eq.~\eqref{prob:opt_P_num}.
We show the optimal MCF $\BH(\theta_{1},\theta_{2})$ in Fig.~\ref{fig:fig3}.
It is notable that we observe three oblique lines, $\theta_2 = \theta_1$ and $\theta_2 = \theta_1 \pm \pi$, which correspond to the in-phase and anti-phase synchronized states, respectively. Both components $H_x$ and $H_y$ of the MCF sharply change their signs as these lines are crossed to stabilize in-phase synchronization and destabilize anti-phase synchronization, respectively.

\begin{figure}
\centering
\includegraphics[width=0.48\textwidth]{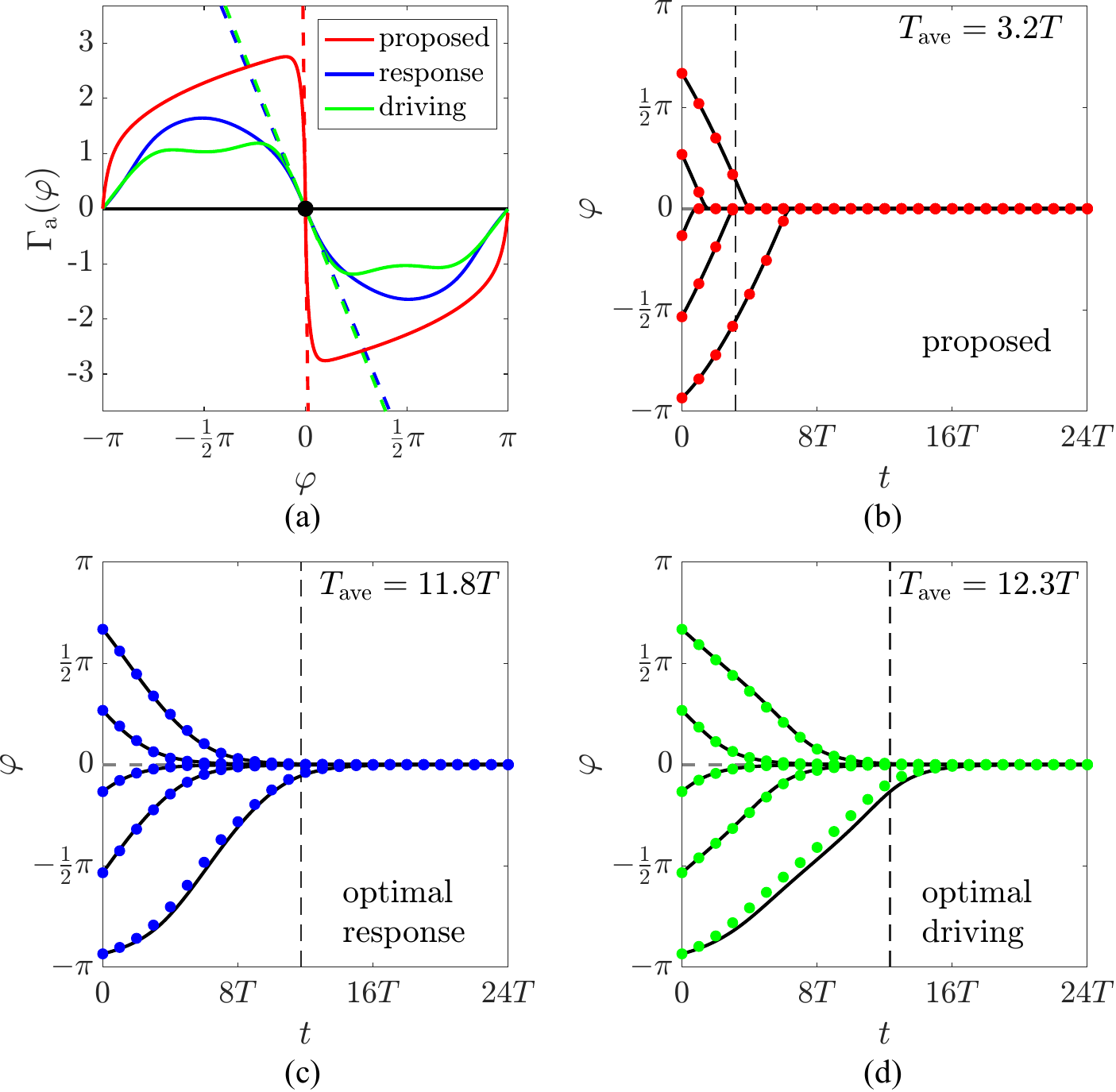}
\caption{
(a)~Optimized APCFs for the coupled identical FHN oscillators.
The red line shows the APCF by the proposed method, the blue line shows the APCF with the optimal response matrix, and the green line shows the APCF with the optimal driving function, respectively.
Each dashed straight line shows the slope of the APCF characterizing the linear stability of in-phase synchronization.
(b)-(d)~Dynamics of the phase difference between the FHN oscillators with the optimized APCFs;
(b)~proposed method, (c)~optimal response matrix, and (d)~optimal driving function.
In (b)-(d), black solid lines represent the convergence of the phase difference $\varphi$ by Eq.~\eqref{eq:phase_diff_sym} and colored dots represent the phase difference~$\varphi$ at $t = \ell T\; (\ell = 0,1,\dots$) obtained by direct numerical simulations of Eq.~\eqref{eq:model}; 
the vertical gray dashed line shows $t = T_{\mathrm{ave}}$ in each figure.
}
\label{fig:fig4}
\end{figure}

We compare the APCFs obtained by different optimization methods as shown in Fig.~\ref{fig:fig4}(a).
For the drive-response optimization, when we optimize the response matrix, the driving function was assumed to be the raw oscillator state, i.e., $\BG(\psi) = \BX_{0}(\psi)$, and when we optimize the driving function, the response matrix was assumed to be a unit matrix, $\BA(\psi) = \mathrm{diag}(1,1)$ (see Appendix~A).
We can confirm that the APCF obtained by the proposed method achieves higher linear stability at $\varphi^{*} = 0$ than the other two drive-response methods.
Indeed, the APCF obtained by the proposed method can achieve arbitrary high linear stability, in principle, because the linear stability can be improved arbitrarily by reducing the weight parameter $\gamma$ as explained in Appendix~B.

We performed numerical simulations of the oscillators given by Eq.~\eqref{eq:model} from five different initial conditions
and compared to the phase equation~\eqref{eq:phase_diff_sym} with the optimal MCF for each method, assuming that the phase reduction and averaging approximation are valid under $\varepsilon = 10^{-2}$. 
The numerical integration was performed by the fourth-order Runge-Kutta~(RK4) method with a fixed time step $\Delta_{t} = T/2M$.
When we directly simulated the states $\BX_{1,2}$ of the oscillators, we regarded the functional $\functional{\tilde{\BH}_{1,2}}{\func{\BX_{1}}{t}(\cdot)}{\func{\BX_{2}}{t}(\cdot)}$ as the function $\functional{\tilde{\BH}_{1,2}}{\func{\BX_{0}}{\theta_{1}(t)}}{\func{\BX_{0}}{\theta_{2}(t)}} = \BH(\theta_{1},\theta_{2})$ of the oscillator phases under phase reduction, 
where $\BH(\theta_{1},\theta_{2})$ was calculated using the phase values evaluated from the oscillator states at each time step.

Figure~\ref{fig:fig4}(b) shows the dynamics of the phase difference $\varphi$ with the MCF obtained by the proposed method,
and Figs.~\ref{fig:fig4}(c) and (d) show the results with the MCF obtained by the previous drive-response methods, where the response matrix is optimized in (c) and the driving function is optimized in (d), respectively.
We can confirm that the proposed method achieves faster convergence to the in-phase synchronized state.
It is remarkable that the convergence time can be reduced to about $1/4$ of that of the previous drive-response method.

\subsubsection{Nearly identical oscillators}
\label{sec:results_2_2}

We next show the results of optimization for a pair of weakly coupled FHN oscillators with slightly nonidentical properties.
The parameters of the vector fields $\BF_{1}$ and $\BF_{2}$ of the oscillators are $(a, b, c) = (1/3, 0.25, 0.16)$ and $(1/3, 0.25, 0.14)$, respectively.
The time scale determined by the parameter $c$ is slightly different from each other, resulting in the frequency mismatch $\Delta = 2.6549$.
We regard the vector field with $(a, b, c) = (1/3, 0.25, 0.15)$ as the common part $\BF$ and the differences of $\BF_{1,2}$ from $\BF$ as $\varepsilon \Bf_{1,2}$, respectively.

\begin{figure}
\centering
\includegraphics[width=0.48\textwidth]{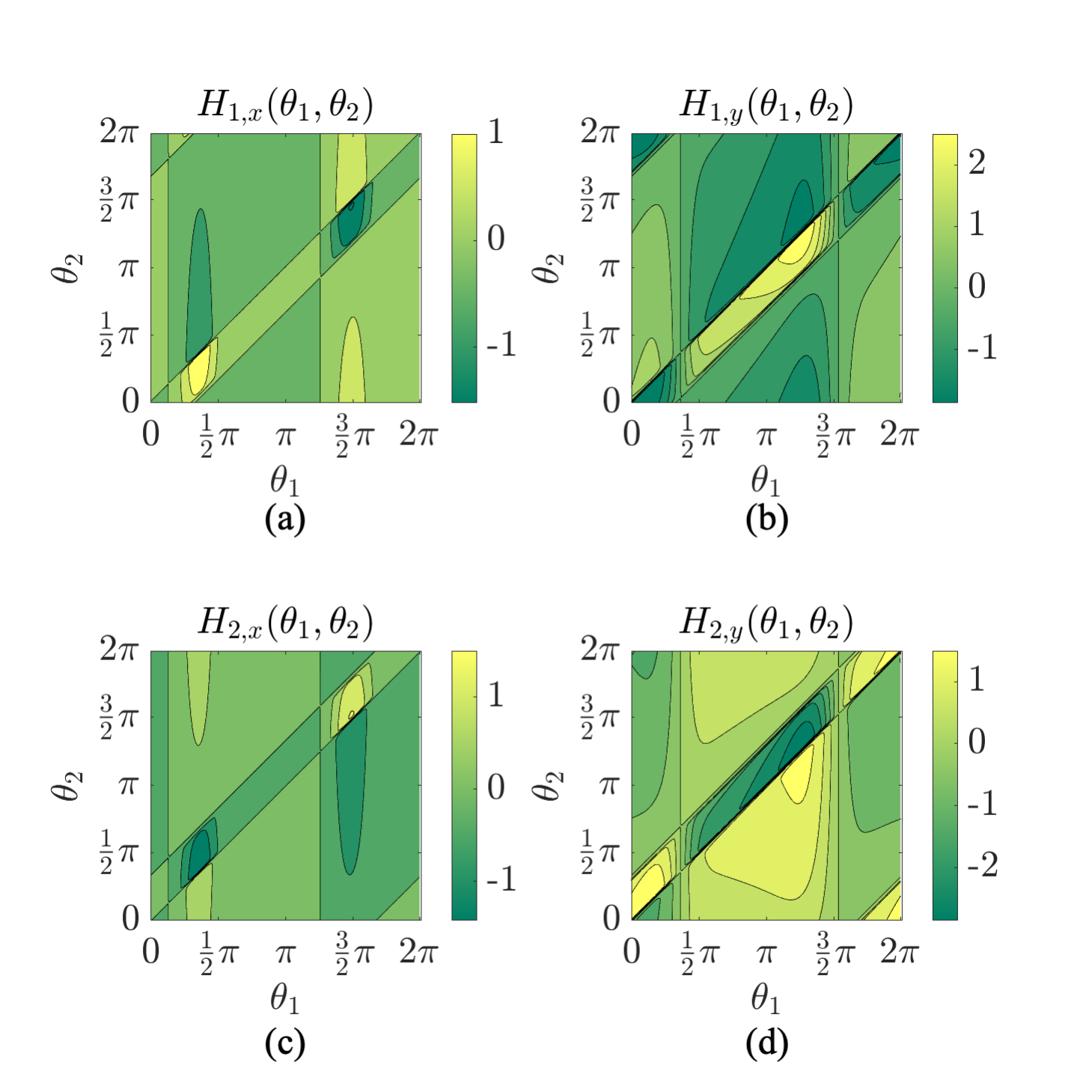}
\caption{
Optimal MCFs $\BH_{1}(\theta_{1},\theta_{2}) = \left[ H_{1,x}(\theta_{1},\theta_{2})\; H_{1,y}(\theta_{1},\theta_{2}) \right]^{\top}$ and 
$\BH_{2}(\theta_{1},\theta_{2}) = \left[ H_{2,x}(\theta_{1},\theta_{2})\; H_{2,y}(\theta_{1},\theta_{2}) \right]^{\top}$ for the coupled nonidentical FHN oscillators.
(a)~$x$ component of $\BH_{1}$.
(b)~$y$ component of $\BH_{1}$.
(c)~$x$ component of $\BH_{2}$.
(d)~$y$ component of $\BH_{2}$.
In each figure, the straight lines represent $\BH_{i,n}(\theta_{1},\theta_{2}) = 0$ for $i = 1,2$ and $n = x,y$.
}
\label{fig:fig5}
\end{figure}

\begin{figure}
\centering
\includegraphics[width=0.48\textwidth]{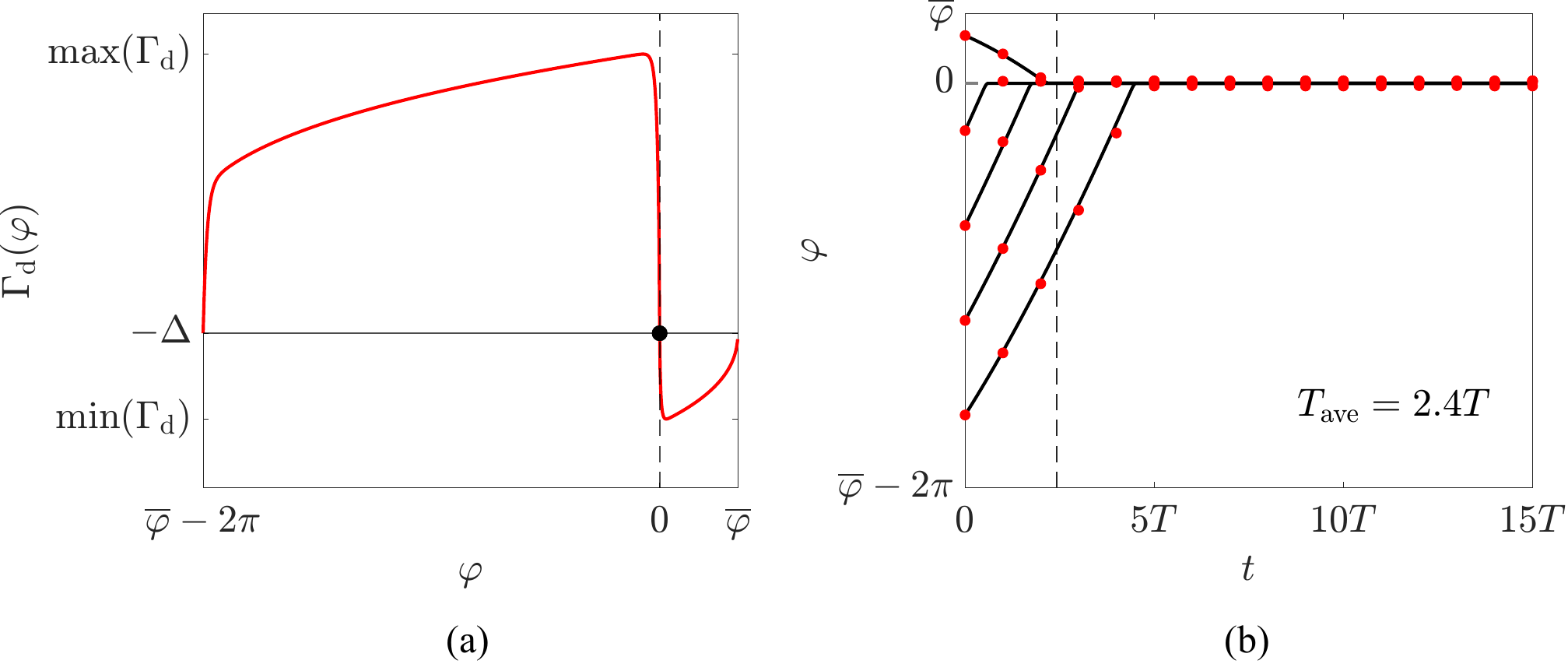}
\caption{
(a)~Optimized DPCF for the pair of nearly identical FHN oscillators.
(b)~Dynamics of the phase difference between the oscillators.
In (b), the solid lines represent the convergence of the phase difference $\varphi$ by Eq.~\eqref{eq:phase_diff} and 
the dots represent the phase difference~$\varphi$ at $t = \ell T\; (\ell = 0,1,\dots$), obtained from the direct numerical simulation of Eq.~\eqref{eq:model}.
The vertical gray dashed line shows $t = T_{\mathrm{ave}}$. 
}
\label{fig:fig6}
\end{figure}

We obtained the optimal MCF with $Q = 1$, $K = 1200$, and $\gamma = 10^{-5}$, assuming $\varepsilon_{\overline{\varphi}} = \varepsilon_{\varphi^{*}} = 2\pi/K$.
The optimal $k^{*}$ was obtained as $k_{\mathrm{opt}}^{*} = 1025$ by a ternary search, corresponding to the unstable fixed point $\overline{\varphi} = 0.9215$ (see Fig.~\ref{fig:fig14}(a) in Appendix).
The optimal MCF~$\BH_{1}(\theta_{1},\theta_{2})$ is shown in Figs.~\ref{fig:fig5}(a) and (b)
and $\BH_{2}(\theta_{1},\theta_{2})$ is shown in Figs.~\ref{fig:fig5}(c) and (d), respectively.
The optimal DPCF $\Gamma_{\mathrm{d}}(\varphi)$ is shown in Fig.~\ref{fig:fig6}(a).
It is interesting to note that $\overline{\varphi}$ is considerably smaller and closer to $0$ than the midpoint $\pi$, 
which indicates that it is easier for the phase difference $\varphi$ to converge to $\varphi^{*} = 0$ from the negative side ($\varphi < 0$) than from the positive side ($\varphi > 0$).
This is due to the difference in the natural frequency of the two oscillators.

We performed numerical simulations of the oscillators given by Eq.~\eqref{eq:model} 
with $\varepsilon = 10^{-2}$ and $\Delta_{t} = T/K$ from five different initial conditions
and compared with the phase equation~\eqref{eq:phase_diff}.
The dynamics of the phase difference $\varphi$ by the proposed method is shown in Fig.~\ref{fig:fig6}(b).
The average convergence time is only about $2.4$ times the period, which is as fast as the case of identical oscillators. 
Thus, the proposed method can be used to synchronize oscillators efficiently even if their properties are slightly different.

\subsection{R\"{o}ssler oscillator}
\label{sec:results_3}

\subsubsection{Symmetrically coupled identical oscillators}
\label{sec:results_3_1}

\begin{figure}
\centering
\includegraphics[width=0.48\textwidth]{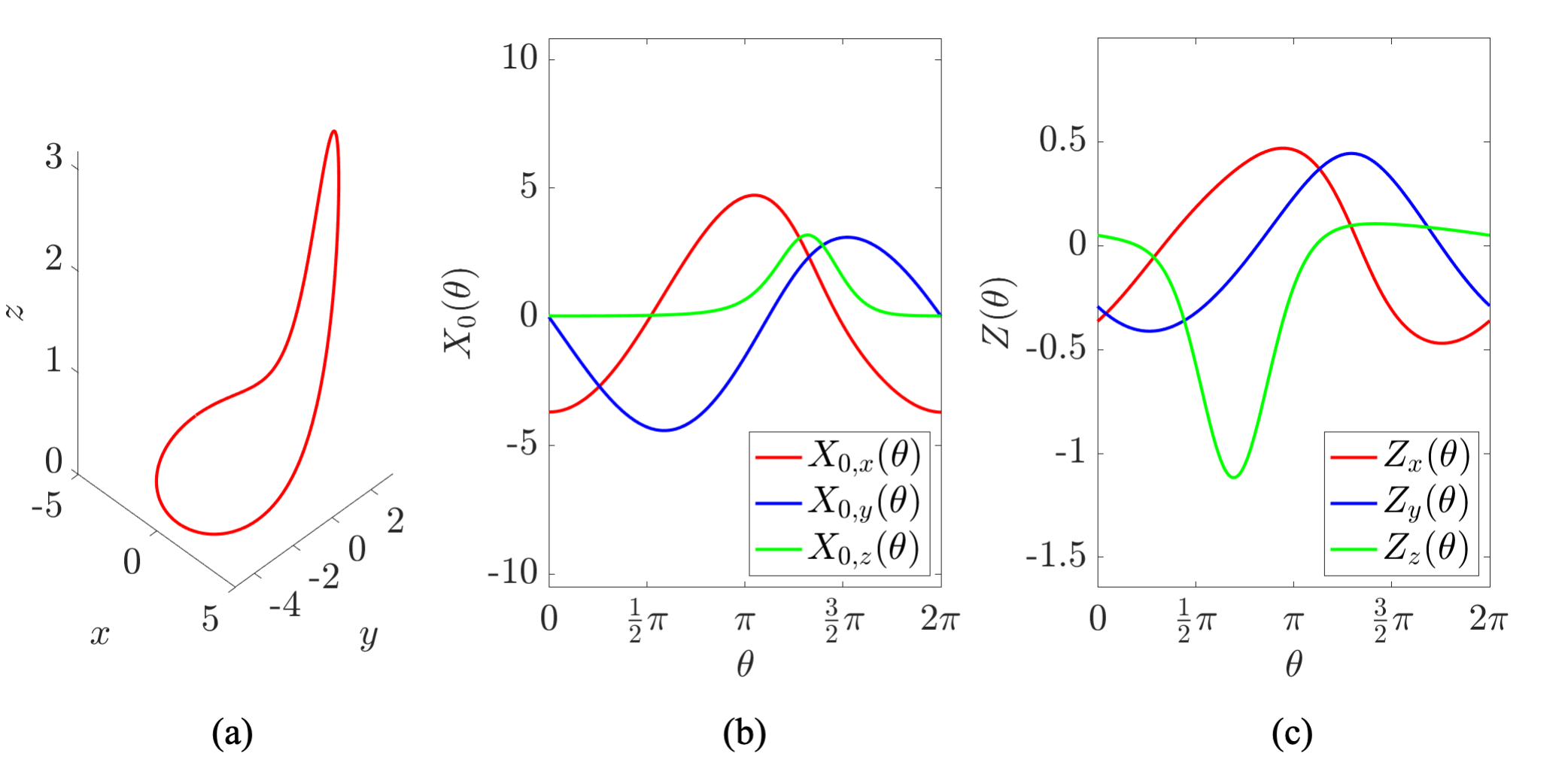}
\caption{
Limit cycle and PSF of the R\"{o}ssler oscillator.
(a)~Limit cycle in the $xyz$ space. 
(b)~Limit cycle solution $\BX = [X_{0,x}\; X_{0,y}\; X_{0,z}]^{\top}$ vs. phase $\theta$. 
(c)~PSF $\BZ = [Z_{x}\; Z_{y}\; Z_{z}]^{\top}$ vs. phase $\theta$.
}
\label{fig:fig7}
\end{figure}

As the second example, we consider a pair of weakly and symmetrically coupled R\"{o}ssler oscillators~\cite{Rossler1976equation} with identical properties, described by Eq.~\eqref{eq:model} with $\BF_{1,2} = \BF$ and $\tilde{\BH}_{1,2} = \tilde{\BH}$.
The R\"{o}ssler oscillator is represented as 
\begin{align}
\dot{\BX} = 
\begin{bmatrix}
\dot{x} \\ \dot{y} \\ \dot{z}
\end{bmatrix}
= 
\begin{bmatrix}
-y - z \\
x + py \\
q + z(x - r)
\end{bmatrix}
,
\end{align}
where we assume $(p, q, r) = (0.2, 0.2, 2.5)$.
This oscillator has a limit cycle with a period $T = 5.745$ and natural frequency $\omega = 1.0937$.
We show the limit cycle in the $xyz$ space in Fig.~\ref{fig:fig7}(a), limit-cycle solution as a function of the phase in Fig.~\ref{fig:fig7}(b), and PSF calculated from the adjoint equation in Fig.~\ref{fig:fig7}(c), respectively.

\begin{figure}
\centering
\includegraphics[width=0.3\textwidth]{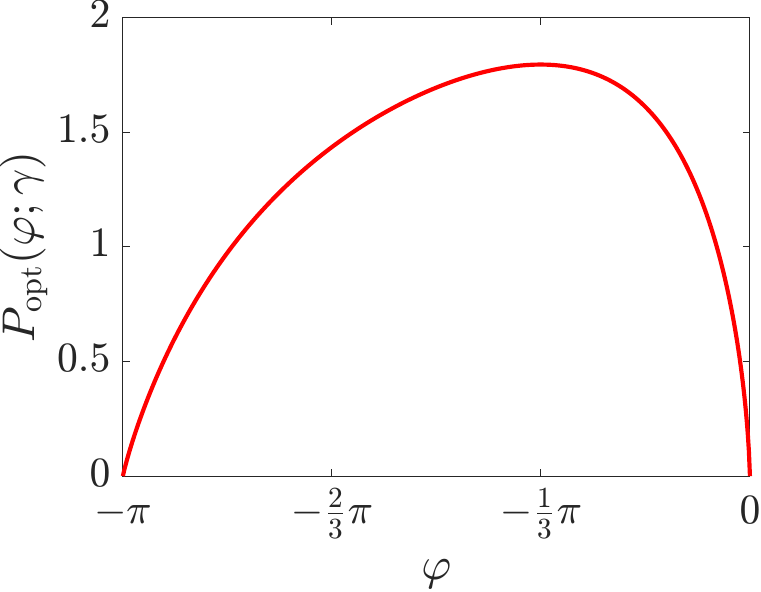}
\caption{Functional form of the optimal amplitude $P_{\mathrm{opt}}(\varphi;\gamma)$ for coupled R\"{o}ssler oscillators ($Q = 2$, $M = 600$, and $\gamma = 10^{-1}$).}
\label{fig:fig8}
\end{figure}

\begin{figure*}
\centering
\includegraphics[width=0.9\textwidth]{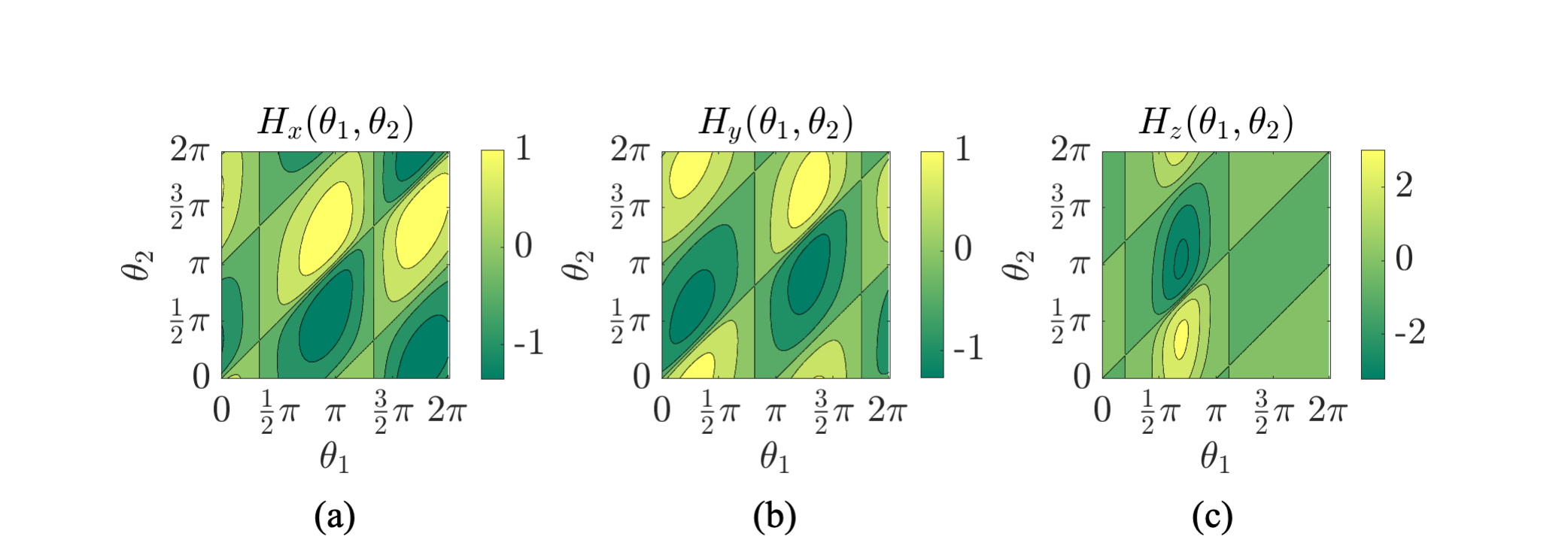}
\caption{
Optimal MCF $\BH(\theta_{1},\theta_{2}) = \left[ H_{x}(\theta_{1},\theta_{2})\; H_{y}(\theta_{1},\theta_{2})\; H_{z}(\theta_{1},\theta_{2}) \right]^{\top}$ for the symmetrically coupled identical R\"{o}ssler oscillators.
(a)~$x$ component {$H_{x}$}.
(b)~$y$ component {$H_{y}$}.
(c)~$z$ component {$H_{z}$}.
In each figure, the straight lines represent $\BH_{x,y,z}(\theta_{1},\theta_{2}) = 0$.
}
\label{fig:fig9}
\end{figure*}

We first show the results by the proposed method.
The optimal amplitude $P_{\mathrm{opt}}(\varphi;\gamma)$ is plotted in Fig.~\ref{fig:fig8}, where $Q = 2$, $M = 600$, and $\gamma = 10^{-1}$, and the small parameters are chosen as $\varepsilon_{\overline{\varphi}} = \varepsilon_{\varphi^{*}} = \pi/M$.
We chose a smoother functional shape of $P_{\mathrm{opt}}(\varphi;\gamma)$ than that of the FHN oscillator to avoid sharp variations near the stable fixed point.
The optimal MCF $\BH(\theta_{1},\theta_{2})$ is shown in Fig.~\ref{fig:fig9}.
As in the case of the FHN oscillator, we observe three oblique lines corresponding to the in-phase and anti-phase synchronized states, and the sign of the MCF changes when these lines are crossed. 

\begin{figure}
\centering
\includegraphics[width=0.48\textwidth]{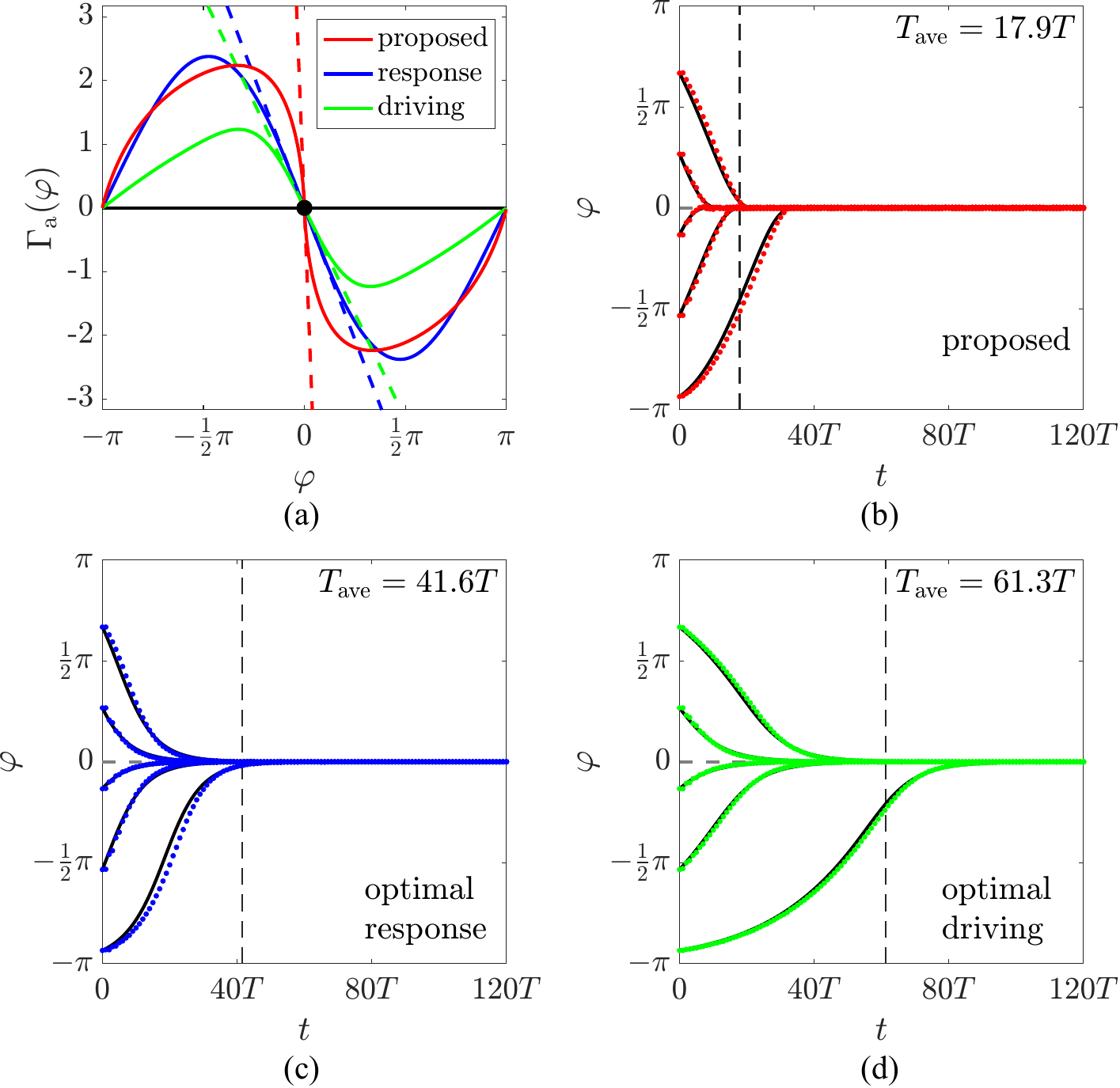}
\caption{
(a)~Optimized APCFs for the coupled identical R\"{o}ssler oscillators.
The red line shows the APCF by the proposed method, the blue line shows the APCF with the optimal response matrix, and the green line shows the APCF with the optimal driving function, respectively.
Each dashed straight line shows the slope of the APCF characterizing the linear stability of in-phase synchronization.
(b)-(d)~Dynamics of the phase difference between the R\"{o}ssler oscillators with the optimized APCFs;
(b)~proposed method, (c)~optimal response matrix, and (d)~optimal driving function.
In (b)-(d), black solid lines represent the convergence of the phase difference $\varphi$ by Eq.~\eqref{eq:phase_diff_sym} and colored dots represent the phase difference~$\varphi$ at $t = \ell T\; (\ell = 0,1,\dots$) obtained by direct numerical simulations of Eq.~\eqref{eq:model}; 
the vertical gray dashed line shows $t = T_{\mathrm{ave}}$ in each figure.
}
\label{fig:fig10}
\end{figure}

We next compare the proposed method to the previous drive-response methods.
As in the case of the FHN oscillator, when optimizing the response matrix, the driving function was given by $\BG(\psi) = \BX_{0}(\psi)$, and when optimizing the driving function, the response matrix was given by $\BA(\psi) = \mathrm{diag}(1,1,1)$.
The optimized APCFs are shown in Fig.~\ref{fig:fig10}(a).
We can again confirm that the APCF by the proposed method achieves higher linear stability at $\varphi^{*} = 0$ than those optimized by the previous drive-response methods.

We performed numerical simulations of coupled R\"{o}ssler oscillators described by Eq.~\eqref{eq:model}
from five different initial conditions using the optimal MCF with $\varepsilon = 10^{-2}$
and compared with the reduced phase equation~\eqref{eq:phase_diff_sym} for each method.
Numerical integration was performed by the RK4 method with a time step $\Delta_{t} = T/2M$.
Evolution of the phase difference $\varphi$ with the APCF by the proposed method is shown in Fig.~\ref{fig:fig10}(b),
and those with the APCFs with the optimal response matrix and with the optimal driving function
are shown in Figs.~\ref{fig:fig10}(c) and (d), respectively.
We can again confirm that the proposed method achieves faster average convergence time than the previous drive-response methods.
It should be noted that the convergence time can be reduced to about $1/3$ of the previous methods in this case.

\subsubsection{Nearly identical oscillators}
\label{sec:results_3_2}

Next, we optimized the MCFs of a pair of weakly coupled R\"{o}ssler oscillators with nearly identical properties.
The parameters of the vector fields $\BF_{1}$ and $\BF_{2}$ are $(p, q, r) = (0.2, 0.2, 2.4)$ and $(0.2, 0.2, 2.6)$, respectively, which yields a frequency mismatch $\Delta = 0.2988$.
We consider a vector field with $(p, q, r) = (0.2, 0.2, 2.5)$ as the common part $\BF$, and defined the phase with respect to this $\BF$.

\begin{figure*}
\centering
\includegraphics[width=0.9\textwidth]{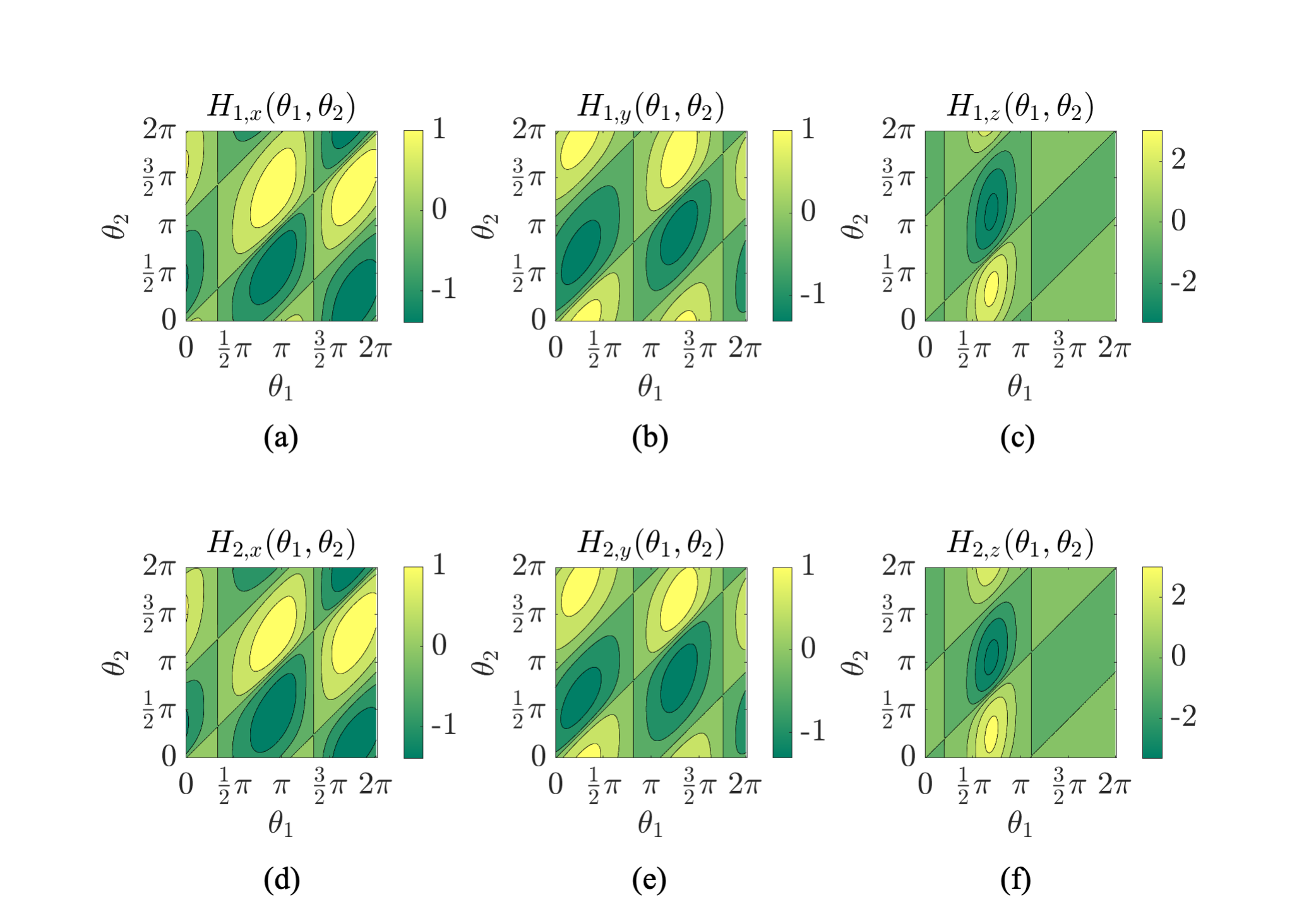}
\caption{
Optimal MCFs $\BH_{1}(\theta_{1},\theta_{2}) = \left[ H_{1,x}(\theta_{1},\theta_{2})\; H_{1,y}(\theta_{1},\theta_{2})\; H_{1,z}(\theta_{1},\theta_{2}) \right]^{\top}$ 
and $\BH_{2}(\theta_{1},\theta_{2}) = \left[ H_{2,x}(\theta_{1},\theta_{2})\; H_{2,y}(\theta_{1},\theta_{2})\; H_{2,z}(\theta_{1},\theta_{2}) \right]^{\top}$ for the coupled nonidentical R\"{o}ssler oscillators.
(a)~$x$ component of $\BH_{1}$.
(b)~$y$ component of $\BH_{1}$.
(c)~$z$ component of $\BH_{1}$.
(d)~$x$ component of $\BH_{2}$.
(e)~$y$ component of $\BH_{2}$.
(f)~$z$ component of $\BH_{2}$.
In each figure, the straight lines represent $\BH_{i,n}(\theta_{1},\theta_{2}) = 0$ for $i = 1,2$ and $n = x,y,z$.
}
\label{fig:fig11}
\end{figure*}

\begin{figure}
\centering
\includegraphics[width=0.48\textwidth]{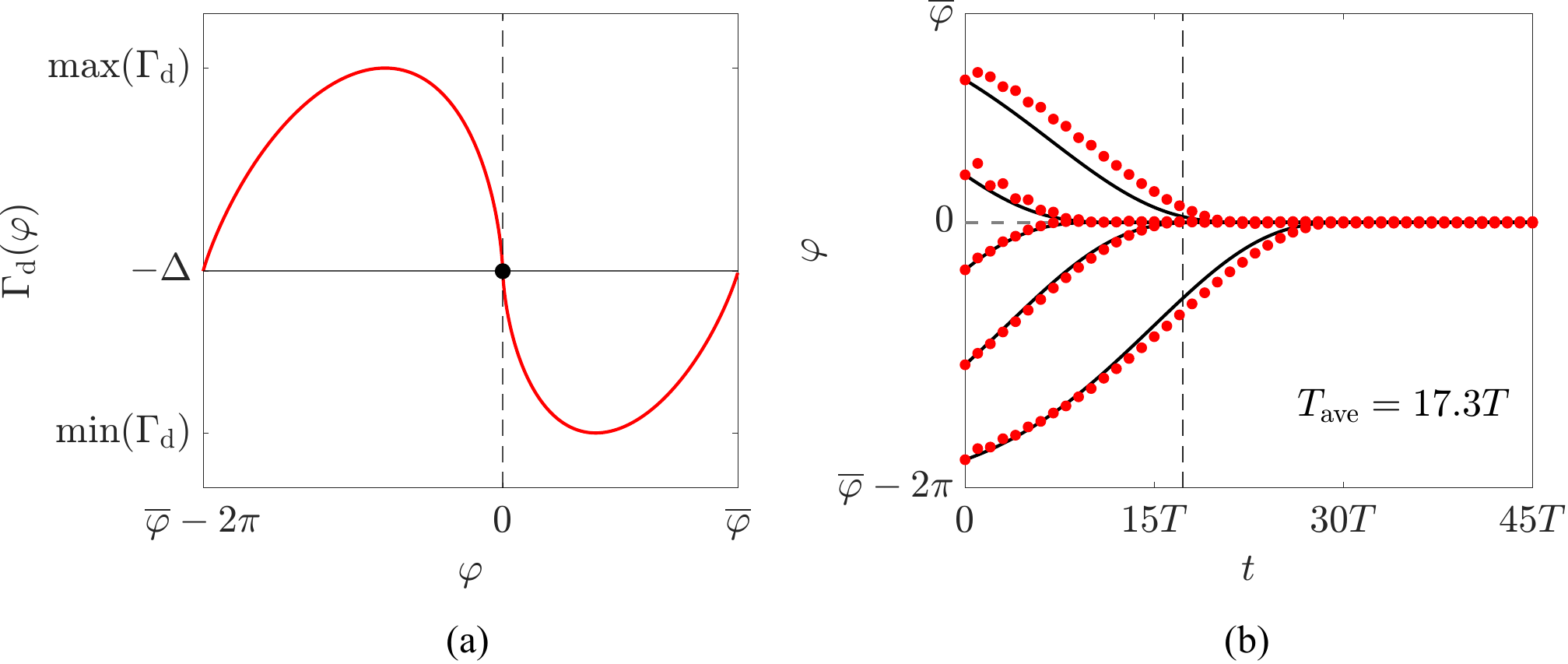}
\caption{
(a)~Optimized DPCF for the coupled nearly identical R\"{o}ssler oscillators.
(b)~Dynamics of the phase difference of the R\"{o}ssler oscillators by the proposed method.
In (b), the solid lines represent the convergence of the phase difference $\varphi$ by Eq.~\eqref{eq:phase_diff} and 
the dots represent the phase difference~$\varphi$ at $t = \ell T\; (\ell = 0,1,\dots$), obtained from the direct numerical simulation of Eq.~\eqref{eq:model}.
The vertical gray dashed line shows $t = T_{\mathrm{ave}}$. 
}
\label{fig:fig12}
\end{figure}

We optimized the MCF for $Q = 2$, $K = 1200$, and $\gamma = 10^{-2}$, where the small parameters were chosen as 
$\varepsilon_{\overline{\varphi}} = \varepsilon_{\varphi^{*}} = 2\pi/K$.
The optimal value of $k^{*}$ was found as $k_{\mathrm{opt}}^{*} = 673$ by a ternary search, which corresponded to $\overline{\varphi} = 2.7646$
(see Fig.~\ref{fig:fig14}(b) in Appendix).
The optimal MCF $\BH_{1}(\theta_{1},\theta_{2})$ is shown in Figs.~\ref{fig:fig11}(a)-(c).
and $\BH_{2}(\theta_{1},\theta_{2})$ is shown in Fig.~\ref{fig:fig11}(d)-(f), respectively.
The resulting optimal DPCF $\Gamma_{\mathrm{d}}(\varphi)$ is shown in Fig.~\ref{fig:fig12}(a).

We performed numerical simulations from five different initial conditions with $\varepsilon = 10^{-2}$ and $\Delta_{t} = T/K$.
The dynamics of the phase difference $\varphi$ with the optimal DPCF obtained by the proposed method is shown in Fig.~\ref{fig:fig12}(b).
The average convergence time is about $17.3$ times the period, which is as fast as the case of identical oscillators.
Due to approximation errors of the phase reduction and averaging, the results of the phase equation~\eqref{eq:phase_diff} slightly differ from the direct numerical simulation of Eq.~\eqref{eq:model}, but they eventually converge to the correct in-phase synchronized state.

\section{Concluding remarks}
\label{sec:conclusion}

In this study, we proposed a method for optimizing the MCFs that lead to fast and global synchronization
for a pair of weakly coupled limit-cycle oscillators with identical or slightly nonidentical properties,
where the coupling between the oscillators can depend not only on their present states but also on their past time series.
We have shown that we can design the optimal coupling function without considering the effect of time delay,
even if the oscillators interact via their past time series,
provided that the reduced phase equations are sufficiently accurate.

The present method does not assume the drive-response structure assumed in the previous study,
which enabled optimization of the MCFs in a wider class of functions.
We revealed that the optimal MCFs by the proposed method can be represented as a product of the PSF
and the amplitude, where the amplitude can be optimized to minimize the convergence time to the synchronized state.
Through numerical simulations, we demonstrated that the oscillators coupled with the optimal MCFs by the
proposed method exhibit significantly faster global synchronization than the previous drive-response methods.

It is quite natural that the optimized MCF is proportional to the PSF, as it obviously provides the most efficient direction
for the driving input to induce the largest instantaneous phase shift of the oscillator. 
The fact that the resulting optimized APCF or DPCF does not depend on the functional form of the PSF indicates that
this indeed offers a universally optimal method to drive the oscillator, irrespective of the detailed characteristics of the oscillator.

We introduced an additional regularization term in the optimization problem for the amplitude of MCF to avoid sharp variations. 
Although the limit with $\gamma \to +0$ theoretically gives the true optimum, it results in a discontinuous change in the APCF or DPCF at $\varphi = 0$, 
leading to strong fluctuations in the dynamics of $\varphi$ near $\varphi = 0$, which hamper stable synchronization. 
In the examples, we empirically chose appropriate values of $\gamma$ and successfully realized fast global synchronization.

In the proposed method of weak mutual coupling, measuring the phase values of the two oscillators is required, which may be an obstacle in practical applications.
In the present setting, the weak coupling between the oscillators is assumed and therefore the oscillator states are always near the limit cycle.
In this case, evaluating the phase values can be relatively easy (see, for example,~\cite{Ota2011measurement,Cestnik2018inferring,Namura2022estimating}).
We could also reduce the number of measurements because the phase of each oscillator increases mostly with a constant frequency and deviations from it are generally small from the weak-coupling assumption.
Also, analyzing the effect of time delay in the phase measurement and developing methods to compensate for it will also be a practical future subject.


\begin{acknowledgements}
We acknowledge JSPS, Japan KAKENHI JP22K11919, JP22H00516, and JST, Japan CREST JP-MJCR1913 for financial support.
\end{acknowledgements}

\appendix


\section{Review of the previous methods}
\label{sec:appendix_1}

Here, we briefly review our previous drive-response method for optimizing mutual synchronization of symmetrically coupled oscillators~\cite{Watanabe2019optimization}, which is compared with our present method in the main text.

In Ref.~\cite{Watanabe2019optimization}, the MCF $\BH(\theta_{1},\theta_{2}) \in \mathbb{R}^{N}$ is assumed to be separated into a response matrix $\BA(\theta_{1}) \in \mathbb{R}^{N \times N}$ of the oscillator and a driving function $\BG(\theta_{2}) \in \mathbb{R}^{N}$ of the other oscillator as
\begin{align}
\BH(\theta_{1},\theta_{2}) = \BA(\theta_{1})\BG(\theta_{2}).
\end{align}
Given one of the response or driving function, the other function is optimized to maximize the linear stability at the in-phase synchronized state.
While the MCF is decomposed into the response or driving function and one of them is assumed to be given in the previous study, 
in this study, we put no such assumption on the MCF and derived that the optimal MCF can be factorized into the PSF $\BZ$ and optimal amplitude $P$ as in Eq.~\eqref{eq:H12}.
Moreover, we optimize not the linear stability of the synchronized state but the average convergence time to it. That is, we focus on the global property (average convergence time) rather than the local property (linear stability), because larger linear stability does not necessarily indicate faster global synchronization.
These are the major differences between the two studies.

First, we consider optimizing the response matrix $\BA$, given a driving function $\BG$.
The optimal response matrix $\BA(\psi)$ that maximizes the linear stability of the in-phase synchronized state is obtained as 
\begin{align}
\BA(\psi) &= -\frac{1}{2\lambda} \BZ(\psi) \frac{d}{d\psi}\BG(\psi)^{\top},
\end{align}
where
\begin{align}
\lambda &= -\sqrt{\frac{1}{4Q_{\BA}} \average{\norm{\BZ(\psi) \frac{d}{d\psi}\BG(\psi)^{\top}}^{2}}{\psi}},
\end{align}
by analytically solving the optimization problem
\begin{align}
\begin{aligned}
&\max_{\BA} && -\frac{1}{2} \Gamma'_{\mathrm{a}}(0)\\
&\mathrm{s.t.} && \average{\norm{\BA(\psi)}_{\mathrm{F}}^{2}}{\psi} = Q_{\BA},
\end{aligned}
\end{align}
where $\norm{\cdot}_{\mathrm{F}}$ is the Frobenius norm.

Next, we consider optimizing the driving function $\BG$, given a response matrix $\BA$.
The optimal driving function $\BG(\psi)$ that maximizes the linear stability of the in-phase synchronized state is obtained as 
\begin{align}
\BG(\psi) &= \frac{1}{2\lambda} \frac{d}{d\psi} \left( \BA(\psi)^{\top} \BZ(\psi) \right), 
\end{align}
where
\begin{align}
\lambda &= -\sqrt{\frac{1}{4Q_{\BG}} \average{\norm{\frac{d}{d\psi} \left( \BA(\psi)^{\top} \BZ(\psi) \right)}^{2}}{\psi}},
\end{align}
by analytically solving 
\begin{align}
\begin{aligned}
&\max_{\BG} && -\frac{1}{2} \Gamma'_{\mathrm{a}}(0)\\
&\mathrm{s.t.} && \average{\norm{\BG(\psi)}^{2}}{\psi} = Q_{\BG}.
\end{aligned}
\end{align}

Our numerical results in the main text show that the method proposed in the present study outperforms the previous methods explained above, indicating that the separation of the MCF into the drive and response structure limits the efficiency of synchronization. 

\section{Effect of the regularization term}
\label{sec:appendix_2}

Here, we discuss the effect of the weight parameter $\gamma$ in the regularization term of the objective function in the optimization problem~\eqref{prob:opt_P}.
Although the case without the regularization term, i.e., $\gamma = 0$, would be truly optimal, the resulting functional form of $P_{\mathrm{opt}}(\varphi;\gamma)$ is undesirable 
because it may cause unnecessary sharp variations in $\varphi$ due to the discontinuity at $\varphi = 0$.
We thus added the sum of the square of the gradient with the weight $\gamma$ to the objective function so that $P$ becomes smoother.

\begin{figure}
\centering
\includegraphics[width=0.35\textwidth]{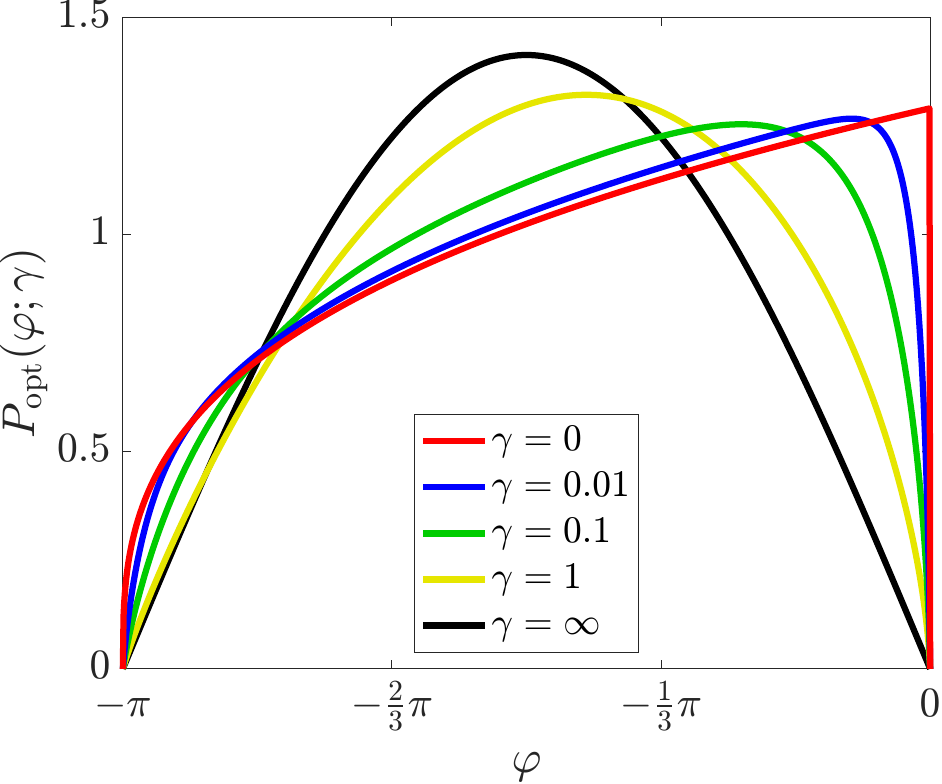}
\caption{
The optimal solutions of $P_{\mathrm{opt}}(\varphi;\gamma)$ for different weight parameters: $\gamma = 0,\; 0.01,\; 0.1,\; 1$, and $\infty$.
The red line shows when $\gamma = 0$,
the blue line shows when $\gamma = 0.01$,
the green line shows when $\gamma = 0.1$,
the yellow line shows when $\gamma = 1$, and
the black line shows when $\gamma = \infty$.
}
\label{fig:fig13}
\end{figure}

We solved this problem for $\gamma = 0,\; 0.01,\; 0.1,\; 1$, and $\infty$ with $Q = 1$ and $M = 600$, where the resulting functional forms of $P_{\mathrm{opt}}(\varphi;\gamma)$ are shown in Fig.~\ref{fig:fig13}.
We can find that the larger the parameter $\gamma$, the smoother the functional form.
We note that $\gamma = \infty$ means the case where only the regularization term of the objective function in Eq.~\eqref{prob:opt_P} is considered 
(i.e., the first term of the objective function is neglected).
In this case, the optimization problem is simply given by
\begin{align}
\begin{aligned}
&\min_{P} && \int_{-\pi}^{0} \left( \frac{dP(\varphi)}{d\varphi} \right)^{2} d\varphi \\
&\mathrm{s.t.} && \frac{1}{\pi}\int_{-\pi}^{0} P(\varphi)^{2} d\varphi = Q, \\
&&& P(\varphi) \geq 0 \quad (\varphi \in [-\pi,0]), \\
&&& P(0) = 0, \\
&&& P(-\pi) = 0.
\end{aligned}
\end{align}
This problem can be analytically solved and the resulting $P(\varphi)$ is proportional to $\sin(\varphi)$ on the interval $[-\pi, 0]$.
In Fig.~\ref{fig:fig13}, we can clearly observe the crossover from the discontinuous function ($\gamma = 0$) to the sine function ($\gamma \to \infty$) of the optimal $P(\varphi)$.
Similarly, in the case of nonidentical oscillators, $\gamma$ also controls the smoothness of the DPCF.

In the practical implementation of the proposed method, 
we should empirically choose an appropriate small hyperparameter $\gamma > 0$ so that the DPCF remains sufficiently smooth near the synchronized state to avoid strong fluctuations in the phase difference.

\section{Numerical method of optimization}
\label{sec:appendix_3}

\begin{figure}
\centering
\includegraphics[width=0.45\textwidth]{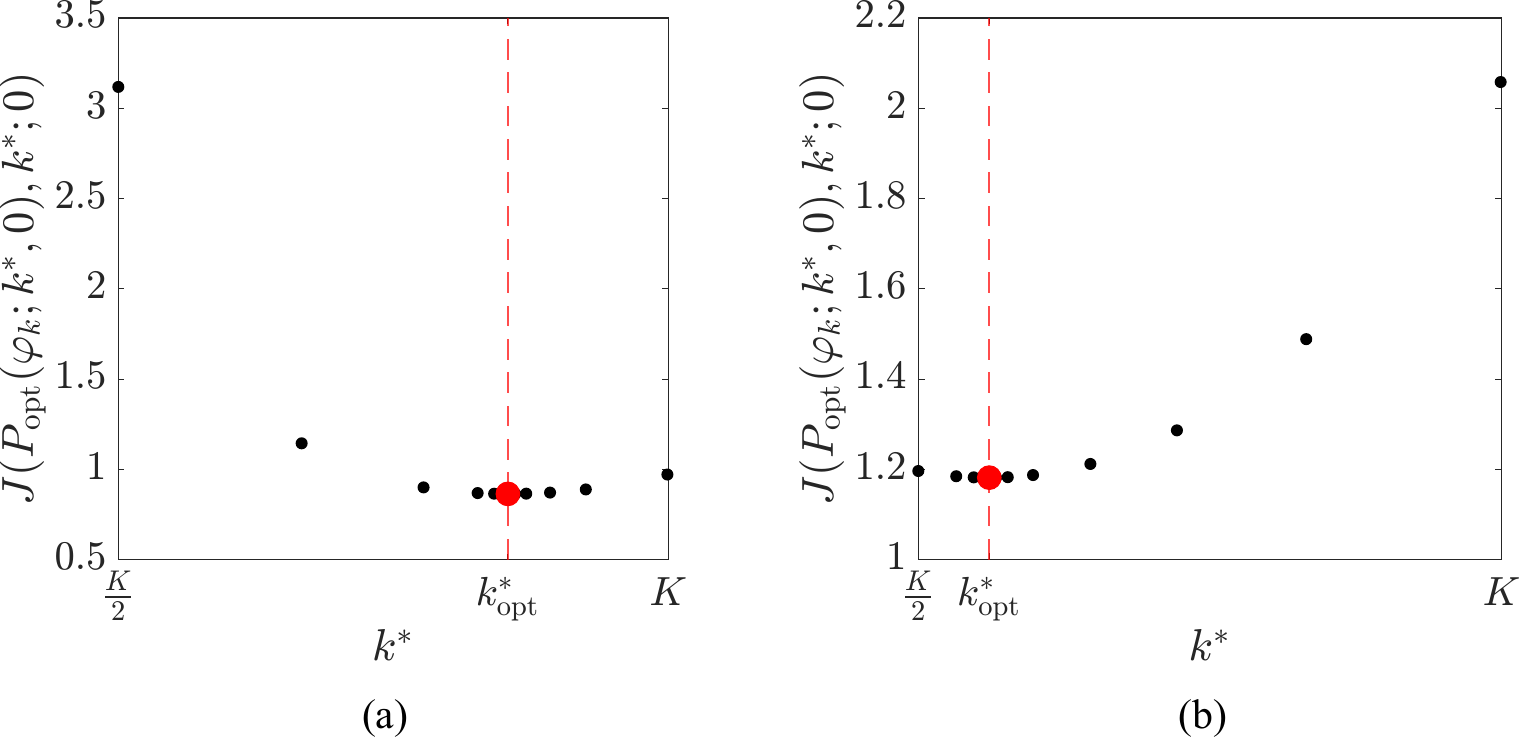}
\caption{
Values of $J(P_{\mathrm{opt}}(\varphi_{k};k^{*},0),k^{*};0)$ and optimal $k_{\mathrm{opt}}^{*}$ by a ternary search.
(a)~FHN oscillators.
(b)~R\"{o}ssler oscillators.
}
\label{fig:fig14}
\end{figure}

Here, we briefly explain the numerical optimization method for the problem~\eqref{prob:opt_P_k_num} in the nonidentical case.
Since it is difficult to optimize $\{ P_{k} \}$ and $k^{*}$ simultaneously, 
we first search the optimal $k_{\mathrm{opt}}^{*}$ with $\gamma = 0$ and then optimize $\{ P_{k} \}$ by using the optimal $k_{\mathrm{opt}}^{*}$ and non-zero $\gamma$.
We denote the objective function of Eq.~\eqref{prob:opt_P_k_num} by $J(P(\varphi_{k}),k^{*};\gamma)$ and the optimal functional form of the amplitude by $P_{\mathrm{opt}}(\varphi_{k};k^{*},\gamma)$, both of which depend on $k^{*}$ and $\gamma$.
The value of the objective function $J(P_{\mathrm{opt}}(\varphi_{k};k^{*},0),k^{*};0)$ with the optimal functional form $P_{\mathrm{opt}}(\varphi_{k};k^{*},0)$ is a function of $k^{*}$ when $\gamma = 0$, but we cannot calculate the derivative of $J(P_{\mathrm{opt}}(\varphi_{k};k^{*},0),k^{*};0)$ with respect to $k^{*}$.
We thus used ternary search for finding the local optimal solution $k_{\mathrm{opt}}^{*}$.
Figure~\ref{fig:fig14} shows the optimal values of $k_{\mathrm{opt}}^{*}$ that minimize $J(P_{\mathrm{opt}}(\varphi_{k};k^{*},0),k^{*};0)$ obtained by ternary search for FHN and R\"{o}ssler oscillators.

\begin{figure}
\centering
\includegraphics[width=0.35\textwidth]{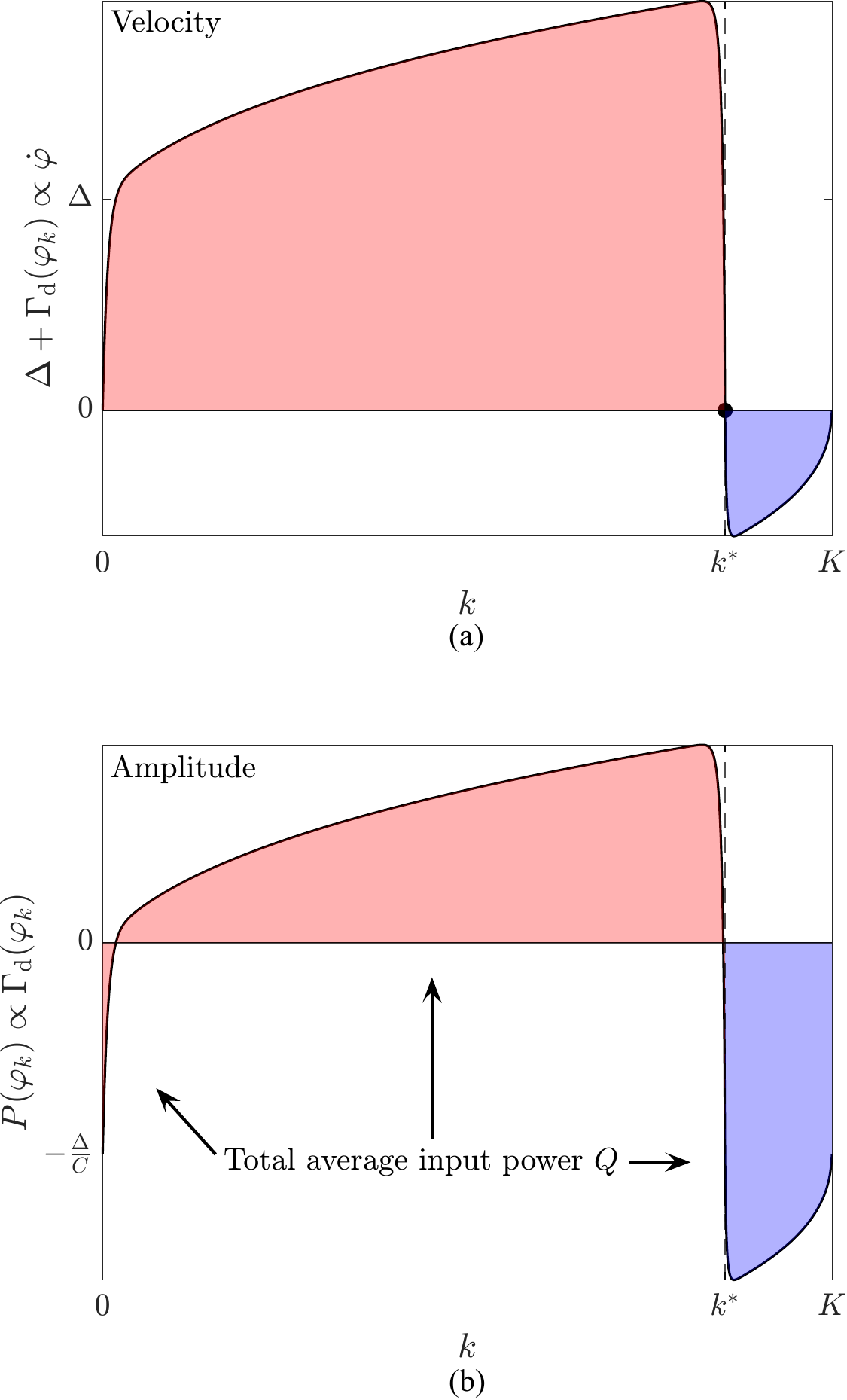}
\caption{
Relationship between the velocity $\dot{\varphi}$ and amplitude $P(\varphi_{k})$ with a total average input power $Q$, where (a) shows the velocity and (b) shows the amplitude for $k \in [0,K]$, respectively, for the case with $\Delta > 0$. 
For $[0, k^{*}]$, the absolute value of the velocity mainly increases with $\Delta$ (red part),
while for $[k^{*}, K]$, the absolute value of the velocity decreases with $\Delta$ (blue part).
Therefore, it is efficient to allocate more power in the interval $[0, k^{*}]$ and not in the interval $[k^{*}, K]$.
}
\label{fig:fig15}
\end{figure}

As shown in Fig.~\ref{fig:fig15},
if $\Delta > 0$, since the velocity is mainly increased with $\Delta$ for $[0, k^{*}]$, while it is decreased for $[k^{*}, K]$, 
it is not efficient to allocate the input power on $[k^{*}, K]$, 
indicating that the interval $[k^{*}, K]$ should be shorter than $[0, k^{*}]$, i.e., $k^{*} > K/2$.
Therefore, we only need to explore the interval $[K/2, K]$.
We note that we cannot choose $k^{*} = K-1,\; K$ because the stable and unstable fixed points are too close.
If $\Delta = 0$, $k^{*} = K/2$ should be optimal from the symmetry of the optimization problem.
Once we find the optimal $k_{\mathrm{opt}}^{*}$, we obtain the optimal $P_{\mathrm{opt}}(\varphi_{k};k_{\mathrm{opt}}^{*},\gamma)$ by introducing an appropriate weight parameter $\gamma$.
After we obtain the optimal $P(\varphi)$, we determine the optimal DPCF $\Gamma_{\mathrm{d}}(\varphi)$ by Eq.~\eqref{eq:Gamma_prop_P}.

Though the global optimum is not completely guaranteed, we generally obtained sufficiently good numerical solutions with the method described here.

\end{document}